\documentclass[]{aastex701}

\usepackage{multirow}
\usepackage{rotating}
\newcommand{\Vcme}{V$_\mathrm{CME}$}

\newcommand{\Frp}{$\mathrm{F_{r,p}}$}
\newcommand{\Fri}{$\mathrm{F_{r,int}}$}
\newcommand{\Fxp}{$\mathrm{F_{X,p}}$}
\newcommand{\tst}{t$_{\rm start}$}
\newcommand{\tpk}{t$_{\rm peak}$}
\newcommand{\trise}{t$_{\rm rise}$}
\newcommand{\dtpk}{${\rm \delta t_{peak}}$}
\newcommand{\Bdur}{${\rm \Delta T_{burst}}$}
\newcommand{\Dcnt}{D$^{90}_{\rm crit}$}
\newcommand{\Dcntf}{D$^{95}_{\rm crit}$}
\newcommand{\kms}{${\rm km\,s^{-1}}$}
\usepackage{amsmath} 

\begin{document}


\title{A comprehensive Sun-as-a-star catalog of interplanetary type-III bursts: Statistics and a framework for stellar exospace weather applications}

\correspondingauthor{Atul Mohan}
\email{atul.mohan@nao.ac.jp}

\author[orcid=0000-0002-1571-7931]{Atul Mohan}
\affiliation{NASA Goddard Space Flight Center, Greenbelt, MD 20771, USA}
\affiliation{The Catholic University of America, Washington, DC 20064, USA}
\affiliation{National Astronomical Observatory of Japan, 2-21-1 Osawa, Mitaka, Tokyo 181-8588, Japan}
\email[]{atul.mohan@nao.ac.jp}

\author[0000-0002-7281-1166]{Sachiko Akiyama}
\affiliation{NASA Goddard Space Flight Center, Greenbelt, MD 20771, USA}
\affiliation{The Catholic University of America, Washington, DC 20064, USA}
\email[]{sachiko.akiyama@nasa.gov}

\author[0000-0002-8182-4559]{Pertti M\"{a}kel\"{a}}
\affiliation{NASA Goddard Space Flight Center, Greenbelt, MD 20771, USA}
\affiliation{The Catholic University of America, Washington, DC 20064, USA}
\email[]{sachiko.akiyama@nasa.gov}

\author[0000-0001-5894-9954]{Natchimuthuk Gopalswamy}
\affiliation{NASA Goddard Space Flight Center, Greenbelt, MD 20771, USA}
\email[]{natchimuthuk.gopalswamy-1@nasa.gov}

\author[0009-0001-3762-5421]{Joshua Budram}
\affiliation{NASA Goddard Space Flight Center, Greenbelt, MD 20771, USA}
\affiliation{The Catholic University of America, Washington, DC 20064, USA}
\affiliation{Prince George’s Community College, Largo, MD, USA}
\affiliation{A. James Clark School of Engineering, The University of Maryland, College Park, MD, USA
}
\email[]{}

\begin{abstract}
Decameter-hectometric (DH) type-III bursts, produced by energetic electron beams escaping into interplanetary (IP) space, are often associated with strong soft X-ray (SXR) flares, coronal mass ejections (CMEs), and solar energetic particle events ({SEPE}s). { With future DH missions being planned, these intense bursts could provide reliable exospace weather event diagnostics, given the lack of direct stellar IP particle and CME probes.} We present a catalog of DH type-III bursts associated with 538 out of 734 $\geq$C5-class flares observed by Wind and STEREO spacecraft during 2006--2025. { Independent of the SXR flare strength, the occurrence and Sun-as-a-star emission properties of DH type-III bursts distinguish confined flares, CMEs with and without strong IP shocks, and SEPEs. We find that logistic-regression models combining SXR and type-III properties provide more accurate and well-calibrated CME/SEPE occurrence probabilities than a model based only on SXR parameters. Further, a space weather event probability-estimation model based on physically meaningful, scale-free parameters derived from SXR and type-III properties outperforms an SXR parameter-only model. 
This result presents a way to define solar-calibrated exospace weather prediction models extendable to stellar flares with a range of energy scales but similar underlying physics.} 
We also present a catalog of 26 shock-associated type-III (S-type-III) bursts that appear to start from cotemporal DH type-II bursts and are generally associated with fast halo CMEs and major { SEPE}s. 
Our catalogs and the scale-free model { form} a benchmark for exospace weather studies.
\end{abstract}



\section{Introduction} 
Type-III bursts are the most common solar radio bursts, appearing as rapidly drifting bright features in dynamic spectra~\citep{wild1970,Mclean85_book,Reid2014}. They are produced by nonthermal electron beams { via coherent plasma emission processes at harmonics} of the local plasma frequency~\citep{ginzburg1958,Tsytovich69,melrose1970,cairns85_secondHarmplasmaemiss}. Metric type-III bursts trace electron beams propagating in the lower corona, while decameter--hectometric (DH; 1\,--\,14\,MHz) and kilometric bursts probe particles escaping into interplanetary (IP) space and are therefore particularly relevant to space weather studies~\citep{gopal04_IPbursts_overview}. 

Flare-reconnection events often produce isolated or groups of type-III bursts lasting several minutes, referred to as complex type-III bursts~\citep{Reiner00_cmplxtypeIII_origin,gopal04_IPbursts_overview,Reid2014}. Type-III storms lasting for $\sim$hours\,--\,days form another class of events associated with interchange reconnection near the boundaries of active regions and open magnetic field structures~\citep{fainberg70_typeIIIstorm,stewart72_typeI-typeIIIstormlink,Aubier78_typeIIIstorm-typeIlink,delzanna11_unifiedMod-typeI-typeIIIstrm}. 
Rarely, a subclass of type-III bursts is found that originates and drifts down from the DH type-II burst features in the dynamic spectrum~\citep{1999GeoRL..26..397R,gopal00_DH-richCMEs-stronger,Cane02_flare-typeIII-CME-SEP-link}. We refer to these shock-associated events as S-type-III bursts, which can diagnose shock-accelerated electrons and { other energetic particles}~\citep[e.g.][]{cane81_herringBtypeII-typeIII-SEPlink,gopal00_DH-richCMEs-stronger,rodriguez25_SEP-cmplxtypeIII}.
\begin{figure*}[ht!]
\epsscale{0.4}
\plotone{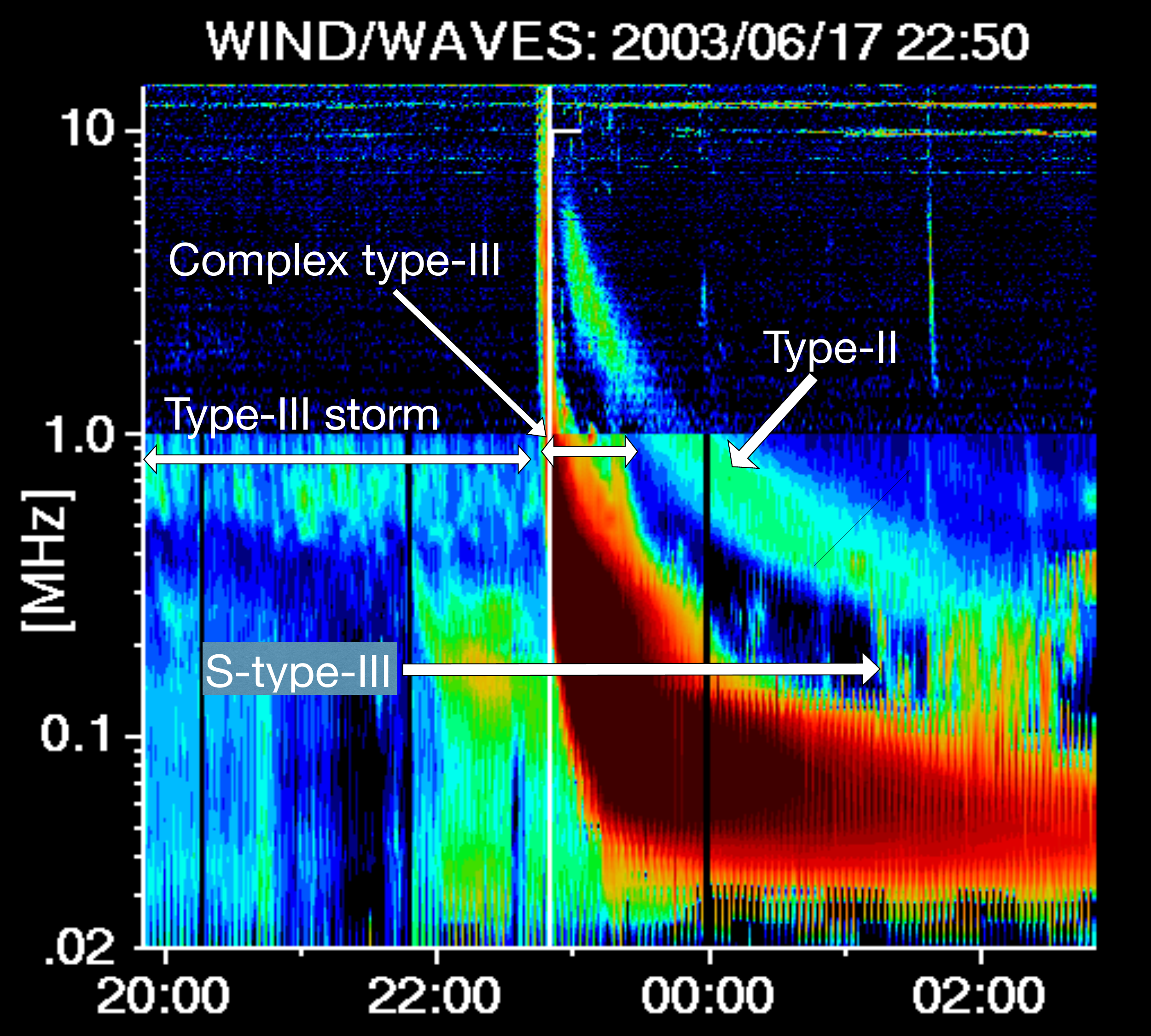}
\caption{Illustration of the various kinds of DH type-III bursts. A type-III storm is seen before the start of a complex type-III burst associated with an M6.8 class flare. The event produced a halo CME which caused the DH type-II and S-type-III bursts.
\label{fig:typeIII_types}}
\vspace{-2mm}
\end{figure*}
Figure~\ref{fig:typeIII_types} illustrates the various kinds of DH type-III bursts during a GOES M6.8 class flare associated with a fast (1813\,\kms) halo CME and a major solar energetic particle event ({SEPE}). A major { SEPE}  is one with $>$10\,pfu proton flux in the $>$10\,MeV band (1\,pfu = 1 particle\,cm$^{-2}$\,s$^{-1}$\,sr$^{-1}$). 
This study focuses on isolated and complex DH type-III and S-type-III bursts.

Owing to their coherent emission mechanism, type-III bursts associated even with weak flares can outshine the solar or stellar disk emission by several orders of magnitude~\citep[e.g.][]{Hilaire,alissa2015, Atul19_typeIIIQPP_turb}, making them excellent Sun-as-a-star diagnostics of energetic electrons and associated space weather events~\citep[e.g.,][]{Lin74_SEErev-typIIIassoc,Miteva22_SEEs-typeIII}. 
Compared to radio burst types in various spectral bands, DH type-III bursts, particularly complex type-III bursts, are more closely associated with major { SEPE}s~\citep{1986AdSpR...6f.319K,Cane02_flare-typeIII-CME-SEP-link,macdowall03_typeIIIl-SEP_link,Miteva17_SEP-radburst_link}.
{ However,} robust relations between DH type-III burst parameters and the occurrence of CMEs or { SEPE}s with various properties have not been found. 
This is likely because major { SEPE}s are produced by CME-driven shocks, whose direct radio diagnostic is a type-II burst~\citep[e.g.][]{gopal05_m-kmTypeII-EnergeticLink}. 
Nevertheless, DH type-II burst events co-temporal with type-III bursts are reported to exhibit particularly high { SEPE} association rates~\citep{Cliver09,Winter15_SEP-DHtypeIII-typeII}. 
Existing studies therefore suggest that combining DH type-III diagnostics with co-temporal radio bursts and { soft X-ray (SXR)} flare properties may improve predictions of { SEPE}, CME, and strong IP shock occurrence and properties~\citep{macdowall03_typeIIIl-SEP_link,gopal23_sgre-gle-ctypeIII}.

In stellar studies, where in-situ measurements and active region imaging are unavailable, coherent radio bursts can provide reliable diagnostics of flare reconnection, CMEs, and { SEPE}s. 
Type-III bursts, being relatively brighter than type-II bursts, could serve as a sensitive stellar space weather diagnostic, which remains to be tested using solar datasets. 
Meanwhile, type-II, III, and IV bursts remain largely elusive in active stellar flare monitoring campaigns~\citep[e.g.][]{2003ApJ...597..535H,villadsen19_VLA_flarestars}, with only recent reports of stellar type-III and type-IV bursts~\citep{atul24_ADLeotypeIV}. This dearth has prompted debates about whether radio-quiet major SXR flares, CMEs, and { SEPE}s could exist on the Sun or whether stellar flares in some active stars follow different emission mechanisms. 
{ Meanwhile, solar observations show that the non-detection of a radio burst during a flare does not necessarily mean that the event is intrinsically radio quiet. Propagation effects and line-of-sight absorption can suppress the emission~\citep[e.g.][]{benz07-radioqietflares, Hillaris16_typeIV,2017A&A...606A.141R, Nasrin18_DHtypeIV_occult_streamerCMEshock}, while coherent emission is also inherently beamed. Using a long-term multi-spacecraft catalog of DH type-IV bursts, \cite{Atul24_DHtypeIVcat} showed that, for any burst event without occulting plasma structures along the lines-of-sight of all simultaneously observing spacecraft, the brightest and most extended event is detected by the spacecraft viewing the active region within $\pm60^\circ$ longitude in its field of view.}

This study aims to build a multi-spacecraft catalog of DH type-III bursts
associated with $\geq$C5-class flares using decades-long observations.
Separate catalogs of flares without type-III bursts and S-type-III bursts
are constructed. We present a statistical analysis of the type-III bursts
and explore their relations with flare and space weather events.
{ We further develop probabilistic models to assess whether DH type-III
properties improve CME and SEPE occurrence prediction beyond SXR flare
properties, and explore a scale-free Sun-as-a-star framework extendable
to stellar flares that vary in physical scales but share common physics.}
Combined with the multi-spacecraft DH type-II
~\citep{Gopal19_DHtypeII_catalog,Atul24_DHtypeIIcat,Atul24_modGBR}
and type-IV~\citep{Atul24_DHtypeIVcat} catalogs, this work forms a
comprehensive Sun-as-a-star database of IP radio bursts and associated
flares, CMEs, and { SEPEs} for solar and stellar space weather studies.
{ 
Our catalogs and statistical analysis could guide solar and stellar space weather assessment using the planned DH-band facilities~\citep[e.g.][]{bentum20_OLFAR,Knapp2024GOLOW,
Deboer25_FARSIDE}.}

Section~\ref{sec:data} describes the data, while Sect.~\ref{sec:methodology} explains the catalog building and analysis methodology. Section~\ref{sec:results} presents the catalog and statistics. Section~\ref{sec:disc} discusses the implications and explores the importance of DH type-III properties in solar and stellar space weather assessment. Our conclusions are presented in Sect.~\ref{sec:conclusion}.
\section{Data} \label{sec:data}
The primary data sources for this work are the SXR data in the 0.5\,--\, 8\AA\ range from GOES, DH dynamic spectra from the Radio and Plasma Wave Investigation \citep[WAVES;][]{WAVES1995} instrument on board Wind, and the Solar TErrestrial RElations Observatory \citep[STEREO;][]{SECCHI2008}
spacecraft. Given the availability of STEREO data, we focus on flares that occurred between December 2006 and January 2025.
We use SXR flare, CME, and major { SEPE} catalogs maintained by NASA's Coordinated Data Analysis Workshops (CDAW) Data Center\footnote{\href{https://cdaw.gsfc.nasa.gov/}{https://cdaw.gsfc.nasa.gov/}}. 
Using these catalogs, we prepared an initial event list of $\geq$C5-class flares that occurred within $\pm$55$^\circ$ longitude on the solar disk. 
We impose this longitude restriction to minimize the impact of line-of-sight effects on burst detectability and to enable reliable magnetic field analysis of flare sites.
For each flare in our initial list, the basic parameters like GOES class, start time (\tst), peak time (\tpk), rise time (\trise=\tpk-\tst), and source (active region) location are recorded along with the properties of associated CMEs and { SEPE}s. 

\section{Methodology}\label{sec:methodology}
DH type-III bursts were searched blindly during all SXR flares in the initial list to enable an unbiased search for radio diagnostics of particle acceleration, CMEs, and major { SEPE}s during every flare. Events missed by this search strategy are mainly those associated with weak SXR flares ($<$C5) or radio bursts with negligible excess SXR emission \citep[e.g.][]{alissa2015,Atul19_typeIIIQPP_turb,Cattel21_typeIII-noSXR}. Individually, such weak events are minor for space weather, though they may contribute collectively to the steady accelerated-particle background~\citep{ash2012_flarestats}, like type-III storms~\citep[e.g.][]{fainberg70_typeIIIstorm,Aubier78_typeIIIstorm-typeIlink,Harra21_AR-typeIIIstorm}. However, we will miss filament eruptions, often associated with weak flares, which can cause SEPE-producing shocks at large IP distances~\citep{2015ApJ...806....8G}.

\begin{figure*}[ht!]
\epsscale{1.15}
\plotone{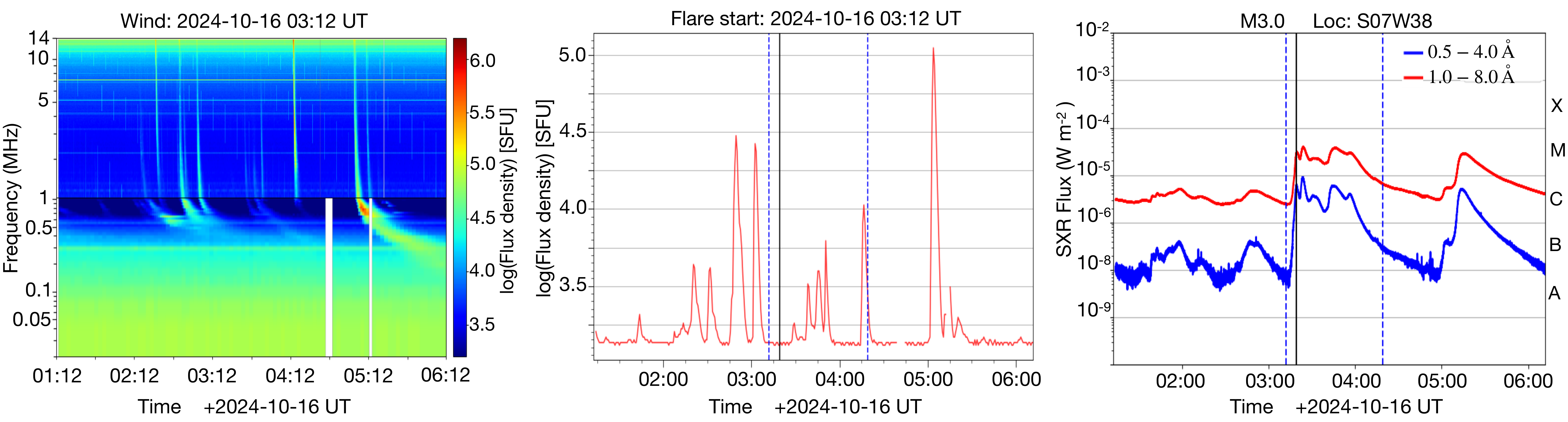}
\caption{A period of overlapping DH bursts and SXR flares. Calibrated DH dynamic spectrum and the 1\,MHz data show multiple type-III spikes, while the SXR light curve shows signs of overlapping flares. The \tst\ and \tpk + 1\,h epochs for the strongest SXR flare during the period are marked by vertical dashed lines, while its \tpk\ epoch is marked by a black line.
\label{fig:complex_ev}}
\vspace{-4mm}
\end{figure*}

{ We retrieved data from the GOES database to generate SXR flare light curves} in the 0.5 – 4 and 1 – 8~\AA\ channels for every event in the initial catalog. Calibrated co-temporal DH dynamic spectra and 1\,MHz light curves were generated for all events, using archival data in the Wind/WAVES and STEREO/WAVES repositories managed by NASA's Space Physics Data Facility (SPDF)\footnote{\url{https://spdf.gsfc.nasa.gov/pub/data/}}. Python codes were developed to automate these tasks.
For each SXR flare, calibrated co-temporal DH dynamic spectrum and the 1\,MHz light curve were manually analysed to identify the associated DH type III bursts between \tst - 0.5\,h  and \tpk + 1\,h. 
If the flare subsides to the background level before \tpk +1\,h, we limit the search window to the flare's end time.
If no burst is identified within the search window, the event is moved to a separate catalog of flares with no/insignificant DH type-III bursts.
If multiple type-III features are present within the search window, the burst that starts within the \tst to \tpk+\trise/2 and shows a distinctive enhanced flux density is chosen. 
This is because a strong type-III burst feature is expected from the standard flare model during this period if open magnetic field structures are associated with the reconnection site~\citep[e.g.][]{hirayam74_flarepromerupt_model_n_geom,shibata95_flarereconn_link,ash14_flareEnergetics_MagE}.
However, periods of overlapping type-III bursts or SXR flares, like in Figure~\ref{fig:complex_ev}, make the SXR–radio association ambiguous. We exclude such periods from the analysis.
For each identified type-III burst, the start and end times, peak flux density (\Frp), and integrated flux density (\Fri) are computed from the 1\,MHz light curve.
The type-III peak time relative to the SXR flare peak time (\dtpk) is recorded. A negative \dtpk\ implies that the radio emission peaks before SXR.
The properties of CMEs (mean projected speed (\Vcme), width), the flux of { SEPE}s, and the occurrence of ground-level enhancements (GLEs) associated with each event are obtained from the catalogs maintained by CDAW~\citep{Yashiro04_LASCOCME_Catalog,Gopal09_LASCO_CMEcatalog,gopal26_LASCOCMEcat_ver2}. 
If no { SEPE} is detected, the flux is reported as -1, while a value of 0 denotes enhanced background activity that prevents a reliable { SEPE} association. For GLEs, we report either 1 to denote an event occurrence or -1 otherwise. 

For every event in the DH type-III catalog, we defined four quality metrics to quantify (i) the confidence in the radio-SXR association (\texttt{Pk-rank}) based on \dtpk, (ii) the reliability of the radio peak flux identification (\texttt{M-score}), (iii) the radio data quality (data\_qlty), and (iv) the confidence in the flare - CME association (\texttt{Ev\_complexity}).
\texttt{Pk-rank} of 1 to 4 were assigned when the radio peak occurs in the intervals (\tst-0.5\,h) to (\tst+\trise/2), (\tst+\trise/2) to (\tpk-$\Delta$t), (\tpk$\pm\Delta$t), and (\tpk+$\Delta$t) to (\tpk+1\,h), respectively, where $\Delta$t= min(3\,min,\trise/2). Larger ranks indicate progressively weaker radio-SXR association.
Type-III bursts can be complex with multiple peaks, or there can be overlapping bursts. \texttt{M-score} rates the confidence in the burst identification and estimation of the peak radio flux. When a single burst with a distinguishable peak can be identified, \texttt{M-score}=1. 
A score of 2 is assigned when reliable radio-SXR event associations can be made, despite the presence of multiple temporally close radio peaks. The peak radio flux density estimates for these events are reliable, but not the integrated flux. 
A \texttt{M-score} of 3 denotes a complex event for which the radio--SXR event association is uncertain, and the peak and integrated flux density estimates are unreliable.
The \texttt{data\_qlty} metric quantifies the radio data quality. A value of 0 denotes poor data with substantial gaps that affected peak and integrated flux density estimates; 1 denotes data with gaps but reliable peak flux estimates, while integrated fluxes may be unreliable; and 2 denotes good-quality data suitable for both peak and integrated flux density measurements.
Finally, the \texttt{Ev\_complexity} takes a value of 0 if the flare and CME identifications are robust, 1 if the CME measurements are uncertain, 2 if the SXR flare profile is impacted by co-temporal emission though \Fxp\ and peak time estimates are robust, and 3 if CME uncertainty and flare profile contamination are present. 

The relationships between DH type-III burst properties and associated space weather events were examined using events with reliable measurements. Additionally, we obtained the DH band data from STEREO A and B spacecraft for every event in the DH type-III catalog and in the catalog of flares without type-III bursts and generated the calibrated dynamic spectra and 1\,MHz light curves. The data from the period when the STEREO spacecraft were significantly apart from Wind offer an opportunity to study the impact of line-of-sight on the observed radio bursts. 

To compile a list of S-type-III bursts, we { visually inspected} all events in the DH type-II catalog\footnote{\url{https://cdaw.gsfc.nasa.gov/CME_list/radio/waves_type2.html}}~\citep{Gopal19_DHtypeII_catalog}, maintained by CDAW, starting from late 1997 to 2025-01-01. We did not apply any selection criteria, including in the source longitude range, to the DH type-II catalog because of the relatively low event count compared to SXR flares.
We assign a `confidence' metric to the detected S-type-III events, where 0 denotes a less confident detection, and 1 denotes a confident detection. The CME and flare characteristics, along with the peak flux of any associated { SEPE}, are tabulated following the same procedure used for the DH type-III catalog. 

\section{Results} \label{sec:results}
We identified 901 $\geq$C5-class flares with co-temporal SXR and DH band data from the GOES and Wind spacecraft, respectively. After removing events that are either complex or have poor-quality data, we obtain a list of 734 events, { of which 311 (42\%) are eruptive ({ CME-associated}), and 36 CMEs are major { SEPE}-associated.} 
{ Of the 734 flares, 538 (73\%) have associated DH type-III bursts, constituting our} DH type-III catalog. 
{ We present an additional catalog of 196 flares without robust type-III emissions, of which 180 (92\%) are confined.}
Meanwhile, among the 675 DH type-II bursts in the CDAW catalog, we found 26 (4\%) events with S-type-III bursts. We provide a catalog of these events with the properties of the associated CMEs, SXR flares, and { SEPE}s. The catalogs are supplemented by calibrated 1\,MHz and SXR light curves, DH dynamic spectra (from the Wind, STEREO A, and STEREO B spacecraft), SEPE flux light curves, CME height-time plots, and links to white light and SXR movies of the Sun. All catalogs and associated data are available online\footnote{\url{https://doi.org/10.5281/zenodo.21133265}}.
Appendix~\ref{app0} presents screenshots of all catalogs.

\subsection{DH type-III catalog}
\begin{figure*}[ht!]
\epsscale{1.05}
\plotone{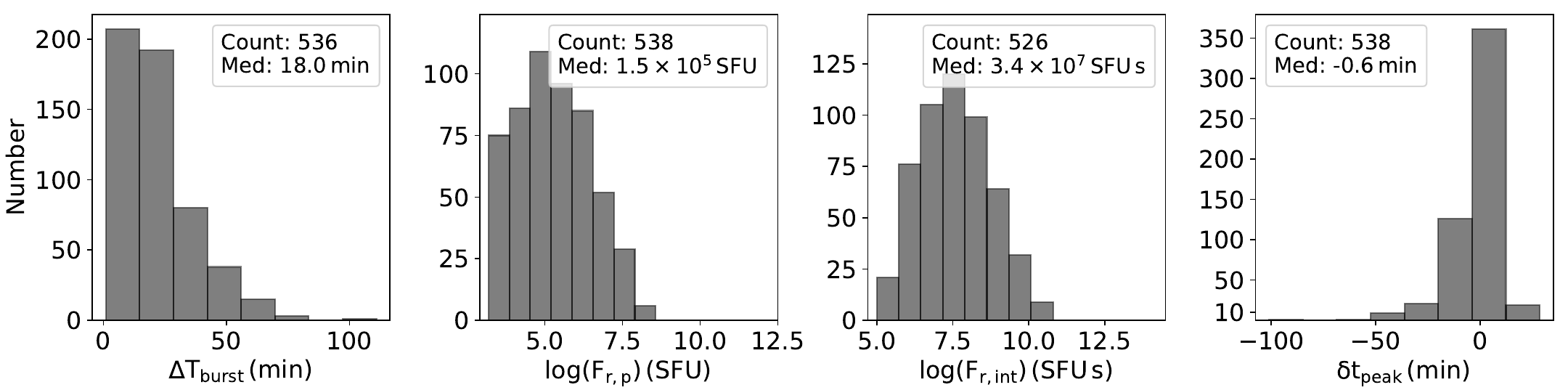}
\vspace{-2mm}
\caption{Properties of all DH type-III bursts in the catalog. Data count and median values are also provided.   
\label{fig:typeIII_stats}}
\vspace{-3mm}
\end{figure*}

Table~\ref{tab:allparams} presents the properties of SXR flares, radio bursts, and associated space weather events, reported in the catalog. {Additionally, the catalog provides the start and end times of the type-III bursts and SXR flares. Figure~\ref{fig:typeIII_stats} shows the histograms and respective median values of the various type-III burst parameters.}

\begin{deluxetable*}{lll}
\tablewidth{0pt}
\tablecaption{Properties of type-III bursts, flares, CMEs, { SEPE}s, and GLE events in the DH type-III catalog. \label{tab:allparams}}
\tablehead{
\colhead{} & \colhead{Property} & \colhead{Description}
}
\startdata
Type-III & $\mathrm{Dur\_min}$ (\Bdur) & Duration in min \\
 & $\mathrm{PeakFlux\_SFU}$ (\Frp) & Peak flux in SFU (Catalog mentions $\log_{10}$\Frp)\\
 & $\mathrm{IntegFlux\_SFUs}$ (\Fri) & Integrated flux in SFU s \\
 & $\mathrm{Time\_wrt\_pflr\_min}$ (\dtpk) & Radio burst peak time relative to the SXR flare peak \vspace{1mm}\\
\hline\vspace{-4mm}\\
SXR flare & $\mathrm{t\_rise\_min}$ (\trise) & Flare rise time in min \\
 & Class & GOES flare class \\
 & FXray (\Fxp) & Peak SXR flux in $\rm W\,m^{-2}$ \vspace{1mm}\\
\hline\vspace{-4mm}\\
CME & Vcme (\Vcme) & Projected CME speed in $\rm km\,s^{-1}$ \\
 & Width & CME angular width in degree \\
 & DHtypeII & Occurrence of DH type II burst, where 1 means yes and 0 means no \vspace{1mm}\\
 \hline\vspace{-4mm}\\
SEPE & $\mathrm{SEP\_max}$ & Peak SEPE flux in pfu \vspace{1mm}\\
\hline\vspace{-4mm}\\
GLE & GLE & GLE occurrence, where 1 means yes and $-1$ means no \\
\enddata
\vspace{-6mm}
\citep{2026IAUS..388..338M}
\end{deluxetable*}

\begin{figure*}[ht!]
\epsscale{1.15}
\plotone{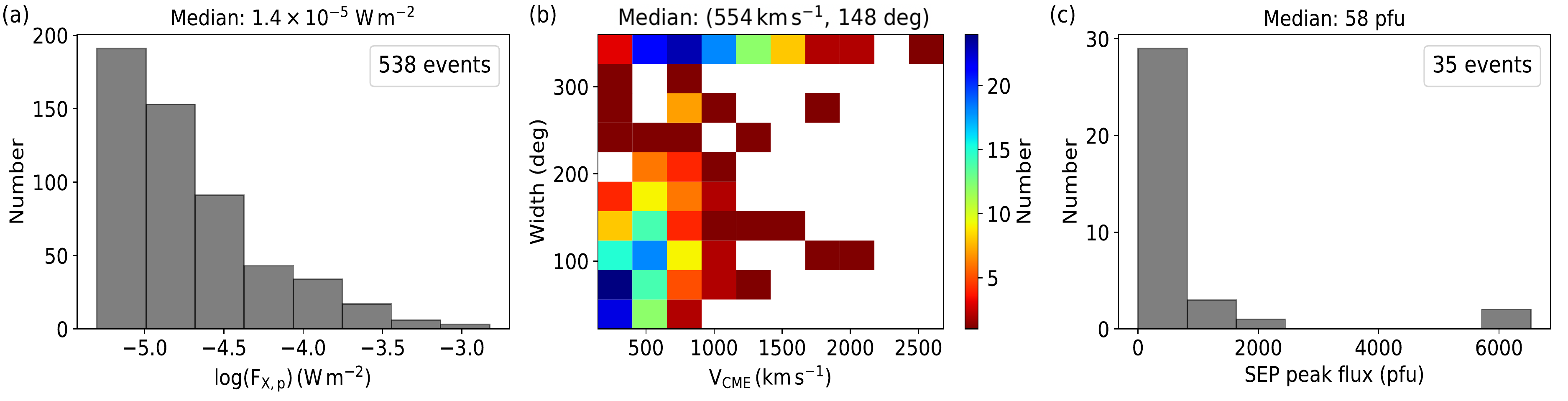}
\vspace{-2mm}
\caption{Properties of all events in the DH type-III catalog. CMEs with reliable measurements were used to make the 2D histogram in panel (b).
\label{fig:event_props}}
\vspace{-4mm}
\end{figure*}

Figure~\ref{fig:event_props} presents the properties of flares, CMEs, and { SEPE}s associated with the events in the DH type-III catalog. The median flare class is M1.4 and the median SEPE peak flux is 58\,pfu. The 2-dimensional (2D) histogram of the CME characteristics reveals { two distinct populations in the \Vcme\,--\,width plane with different median properties: (1) slow-narrow CMEs and (2) fast-halo CMEs.}
\begin{figure*}[ht!]
\epsscale{0.7}
\plotone{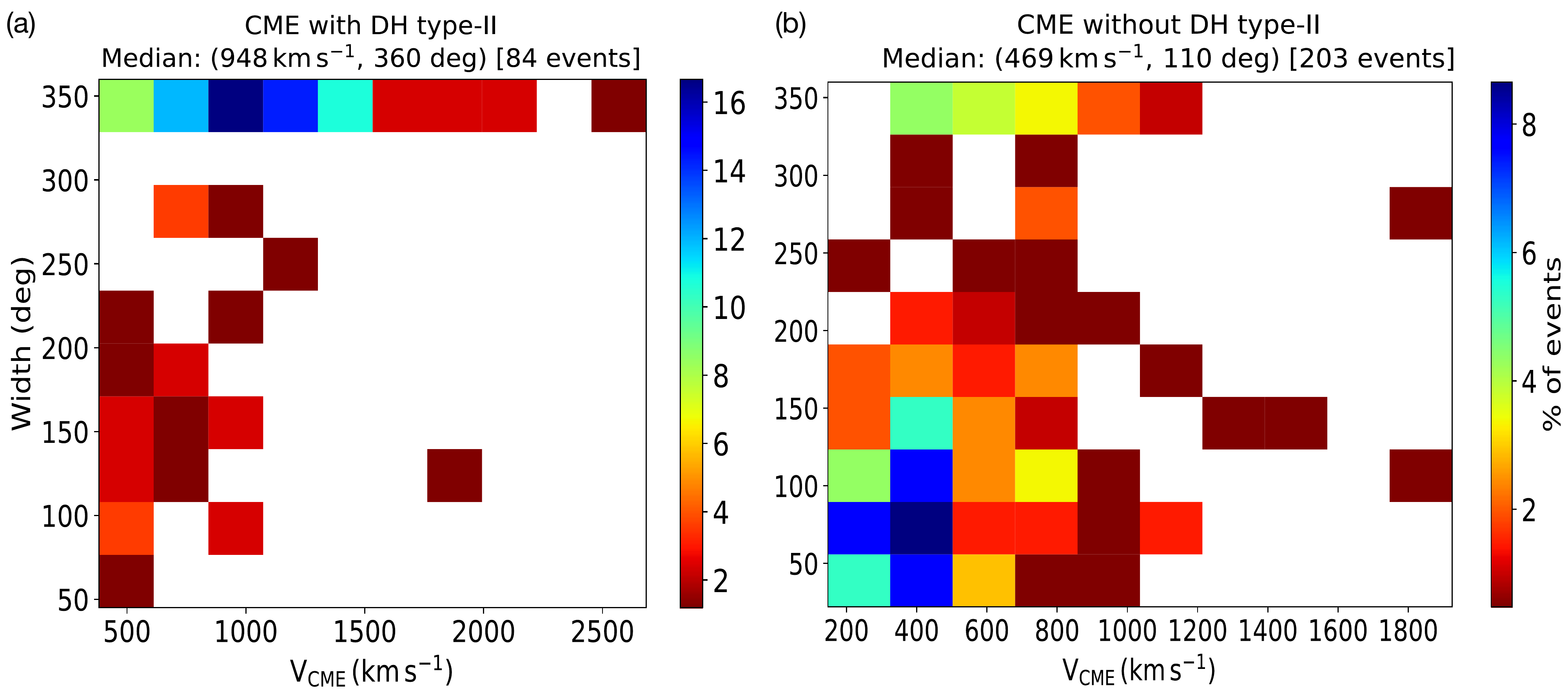}
\vspace{-2mm}
\caption{2D histograms showing the joint distribution of the width and \Vcme\ of CMEs with and without DH type-II bursts in the DH type-III catalog. 
\label{fig:CME_pops}}
\vspace{-4mm}
\end{figure*}
To understand the two populations, Fig.~\ref{fig:CME_pops} { plots the 2D histogram of \Vcme\ and width} for CMEs associated with IP shocks or DH type-II bursts and those without.
Clearly, the { fast-halo population is dominated by} CMEs associated with DH type-II bursts, while the remaining events form a largely slow, non-halo population, consistent with previous studies~\citep{Gopal01_DHII-CMEprops,Gopal08_FWCMEs-noDHII}.
{ For a given coronagraph, halo CMEs are known to be relatively faster with stronger SXR flares than a regular CME~\citep{gopal07_geoeffectiveness_HaloCMEs}. This is because only energetic CMEs expanding fast enough to appear to surround the occulting disk are observed as halos.
However, the direction of CME propagation relative to the observer is a major factor in making a fast CME observable as a halo. 
In our sample, only CMEs originating from active regions within $\pm55^\circ$ longitude are included, excluding events far from the disk center. Hence, \Vcme\ may play a greater role in making a CME observable as a halo. A detailed case-by-case analysis, beyond the scope of this study, would be required to explore this possibility further. We therefore use the occurrence of CME shock-driven DH type-II bursts as a diagnostic of energetic fast-CMEs, rather than a heuristic based on the \Vcme\,--\,width plane.}
\begin{figure*}[ht!]
\epsscale{0.88}
\plotone{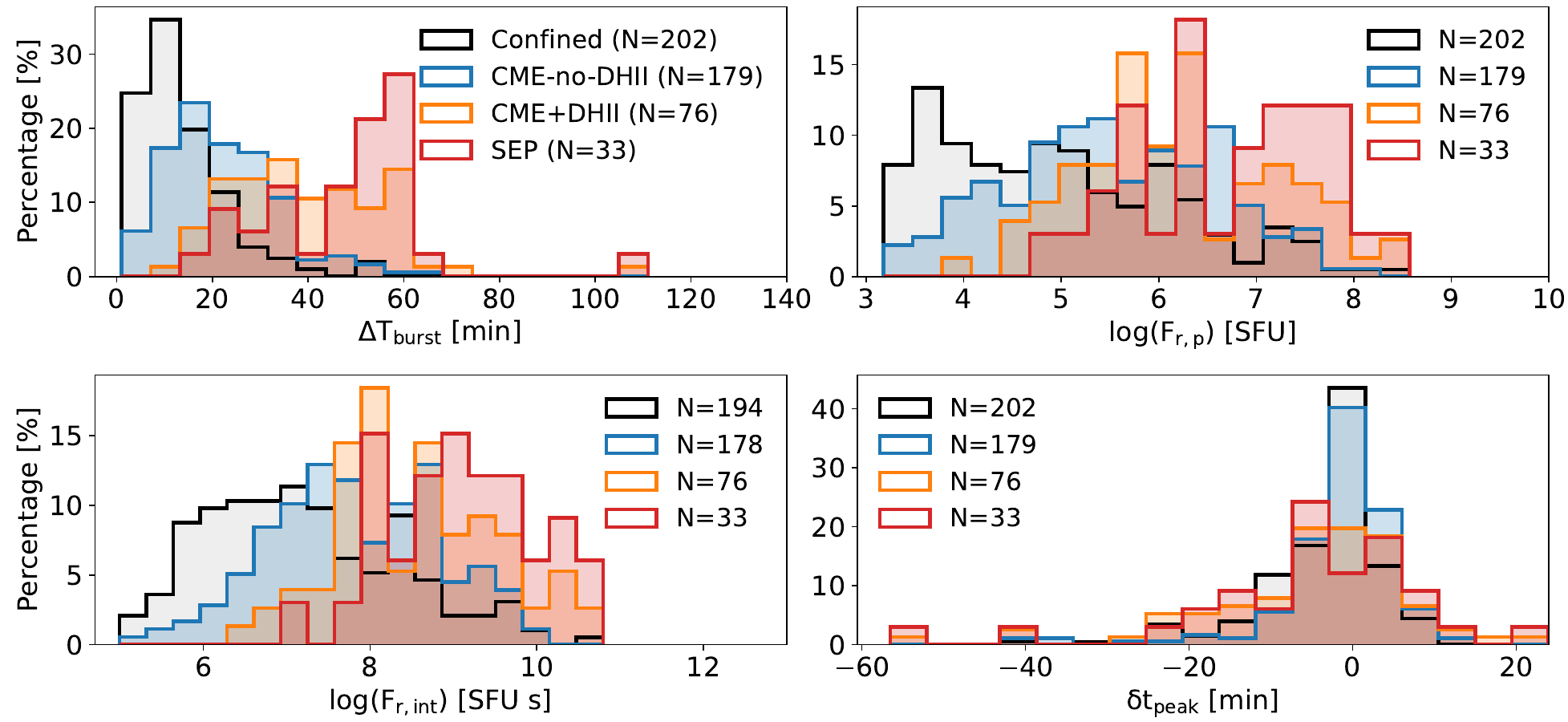}
\vspace{-2mm}
\caption{Histograms of various DH type-III properties for the four different event groups. Distributions for all properties, except \dtpk, show a significant separation in their lower envelope and median value across groups.
\label{fig:grphists}}
\vspace{-3mm}
\end{figure*}

\begin{deluxetable*}{lcccc}
\tablewidth{0pt}
\tablecaption{
Median values (computed in linear space) of the various DH type-III properties for the different event groups.
\label{tab:medprops_grps}
}
\tablehead{
\colhead{Property} &
\colhead{Confined} &
\colhead{CME-no-DHII} &
\colhead{CME+DHII} &
\colhead{SEP}
}
\startdata
\Bdur\, [min] & 12.0 & 21.0 & 37.5 & 51.0 \\
\Frp\, [SFU] & $6.3\times10^{4}$ & $3.5\times10^{5}$ & $1.3\times10^{6}$ & $6.3\times10^{6}$ \\
\Fri\ [SFU s] & $1.4\times10^{7}$ & $5.8\times10^{7}$ & $3.3\times10^{8}$ & $1.2\times10^{9}$ \\
\dtpk\, [min] & $-1.3$ & $-0.2$ & $-2.8$ & $-3.0$ \\
\enddata
\vspace{-5mm}
\end{deluxetable*}
Figure~\ref{fig:grphists} explores the DH type-III properties within four { space weather relevant} event groups, namely: `confined': confined flares, { `CME-no-DHII':} CMEs without DH type-II bursts, { `CME+DHII':} CMEs with DH type-II bursts, and `SEP': { SEPE}s. 
To retain only reliable measurements, the rest of our analysis uses bursts with \texttt{M-score}=1 and starting before the SXR flare \tpk. Although the association of confined flares with type-III bursts may appear surprising, these bursts can arise from jets or minor reconnections involving nearby magnetic structures, while most flare-accelerated particles are confined to the active region~\citep[see,][]{Li19_types_confflares,Chen25_C-flares-typeIII-HXR-noCME}.
Histograms of all burst properties except \dtpk\ show { signs of} separation between event groups, notably in their lower envelopes. The bursts associated with confined flares show the lowest median \Fri\ and \Bdur, indicating that most of the energy in the flare-accelerated electron beams is dissipated close to the flare site. The median values of \Bdur, \Frp, and \Fri\ increase systematically from confined to CME-no-DHII, CME+DHII, and SEP groups (Table~\ref{tab:medprops_grps}). { These inferences from the histogram analysis, together with the finding that 92\% of the SXR flares without associated DH type-III emission are confined, suggest that both the occurrence and properties of flare-associated DH type-III bursts may carry useful space weather diagnostic information.}

\section{Discussion} \label{sec:disc}
This section explores whether the distributions of DH type-III burst properties are statistically distinct across the four event groups and whether these properties, together with flare parameters, can be used to develop { statistical models to estimate the occurrence probabilities of major space weather events}. 
Meanwhile, models based on { extrinsic quantities} like \Fri\ and \Bdur, which depend on Sun-specific physical scales, cannot be generalized to stellar systems. So, we will discuss ways to define { scale-free Sun-as-a-star parameters and associated statistical models applicable to exospace weather studies in Sun-like stars}. 
We further discuss the properties of S-type-III bursts and a small fraction of eruptive flares without { robust DH type-III association in our flare sample}.
\subsection{Properties of DH type-III bursts}\label{sec:DHIII_props}
The histogram analysis in Fig.~\ref{fig:grphists} suggests that the DH type-III properties \Frp, \Fri, and \Bdur\ could distinguish the four event groups. If so, these burst properties { could improve probabilistic models of space weather event occurrence} during $\geq$C5-class flares.
However, we should first analyse the role of the big flare syndrome~\citep{kahler82_BFS} in the results. This is because events with brighter SXR flares typically cause stronger magnetic energy release in the corona and IP space. Consequently, energetic CMEs, stronger IP shocks, and { SEPE}s tend to be associated with stronger SXR flares. If such flares also preferentially produce brighter and longer-duration DH type-III bursts, then the median properties of type-III bursts can appear to increase going across event groups with weaker to stronger magnetic energy release and particle acceleration. Hence, we should test whether the DH type-III burst properties remain statistically distinct across the different event groups even within narrow intervals of SXR flare strength.
\begin{figure*}[ht!]
\epsscale{1.1}
\plotone{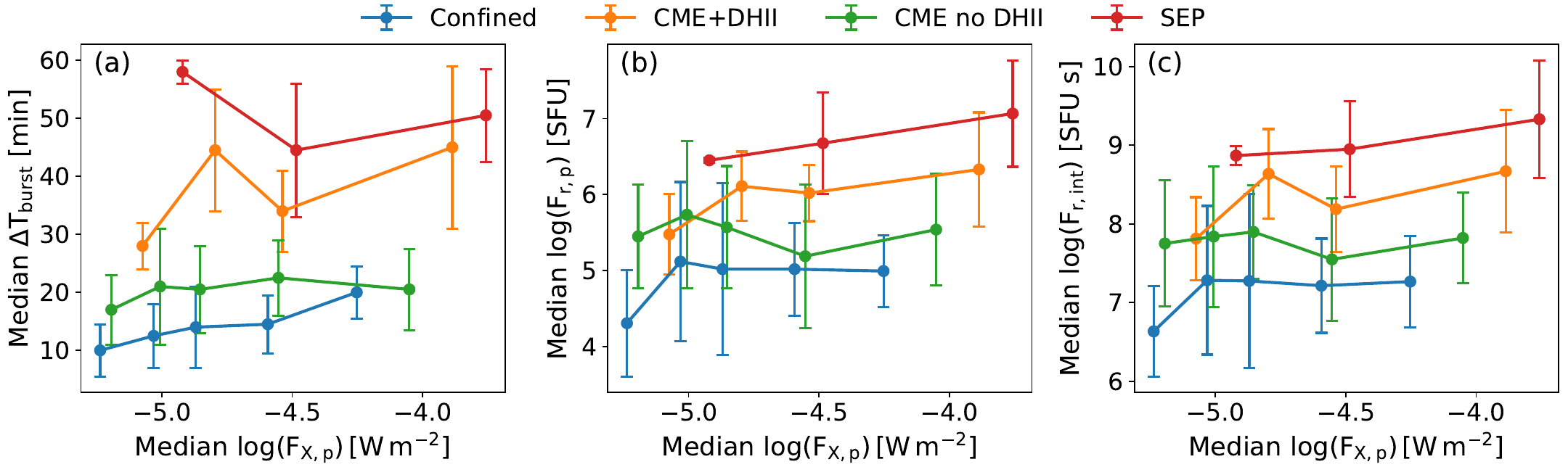}
\vspace{-2mm}
\caption{{ Median \Bdur, \Frp, and \Fri\ values within \Fxp\ bins for the different event groups. Error bars indicate the respective MADs. The median type-III burst properties, particularly \Bdur, show systematic differences among the event groups within most \Fxp\ bins.}
\label{fig_bfs}}
\vspace{-3mm}
\end{figure*}
\subsubsection{Assessing the impact of the big-flare syndrome}\label{sec:bfs}
To assess the impact of the big-flare syndrome, we selected the most reliably detected DH type-III bursts (\texttt{M-score}=1) and divided them into five narrow \Fxp\ bins, each containing $\gtrsim$100 flares. The $\log$\,\Fxp\ ranges of the bins are: (-5.302, -5.097], (-5.097, -4.959], (-4.959, -4.745], (-4.745, -4.387], and (-4.387, -2.824]. 
However, examining the data counts in individual event groups, we find that CME+DHII and SEP groups contain $<$5 events in the $\log$\,\Fxp$<$-4.959 ($<$M1.1) bins. Therefore, to improve statistics, all $<$M1.1 flares in these two groups were merged into a single $\log$\,\Fxp\ bin ((-5.302, -4.959]), while the original \Fxp\ bins were retained for stronger flares. Table~\ref{tab:app1} in Appendix~\ref{app1} provides the median properties of DH type-III bursts for various event groups within each \Fxp\ bin.
Figure~\ref{fig_bfs} shows the median DH type-III burst properties for each event group within every $\log$\,\Fxp\ bin. Error bars show the corresponding median absolute deviations (MADs). The X coordinates indicate the median $\log$\,\Fxp\ values. { Though the MAD intervals overlap, the systematic separation of medians of the type-III burst parameters is evident within many \Fxp\ bins. This separation is very evident in \Bdur, while \Fri\ and \Frp\ can also distinguish the event groups for $>$M1.1 flares. { SEPE}s consistently produce the longest-duration DH type-III bursts with the highest peak and integrated flux densities within a given \Fxp\ bin.} 
Nevertheless, the apparent tendency of median radio flux to increase with \Fxp\ within each group could be due to big-flare syndrome. 

{ The systematic median separation tells us that the central tendency of a property distribution, or the representative property value, shifts between event groups, for a fixed \Fxp\ bin.} 
To assess whether the type-III burst property distributions themselves are statistically distinct across event groups within each \Fxp\ bin, we perform Kolmogorov-Smirnov (K--S) tests using the scipy.stats.ks\_2samp function in SciPy~\citep{Scipy2020}. 
We consider the following event pairs: (confined, CME-no-DHII), (confined, CME+DHII), (CME+DHII, CME-no-DHII), (SEP, CME-no-DHII).
For each type-III property, K-S tests were performed for the selected event pairs within individual $\log$,\Fxp\ bins. The analysis is restricted to $>$M1.1 flare bins, where all event groups contain a sizable number of events. Due to the lack of at least five { SEPE}s in the weaker \Fxp\ bins, the K-S test for the (SEP, CME-no-DHII) pair was 
carried out only in the strongest \Fxp\ bin.
\begin{deluxetable}{llllllllll}[]
\scriptsize
\tablewidth{0pt}
\tablecaption{K-S test results (D and p values) for DH type-III properties in log \Fxp\ bins. Critical D values, D$_{\rm crit}$, are computed for 95\% and 90\% confidence levels using the data counts, N1 and N2.\label{tab:KS_test}}
\tablehead{
\colhead{Property} &
\colhead{$\log({\rm F_{X,p}})$ bin} &
\colhead{Group 1} &
\colhead{$N_1$} &
\colhead{Group 2} &
\colhead{$N_2$} &
\colhead{D} &
\colhead{D$^{95}_{\rm crit}$} &
\colhead{D$^{90}_{\rm crit}$} &
\colhead{p}
}
\startdata
\multirow{10}{*}{\Bdur}
& $(-4.959,-4.745]$ & Confined & 38 & CME-no-DHII & 42 & 0.33 & 0.30 & 0.27 & $2.0\times10^{-2}$ \\
& $(-4.959,-4.745]$ & Confined & 38 & CME+DHII & 10 & 0.74 & 0.48 & 0.43 & $8.0\times10^{-5}$ \\
& $(-4.959,-4.745]$ & CME+DHII & 10 & CME-no-DHII & 42 & 0.60 & 0.48 & 0.43 & $3.0\times10^{-3}$ \\
& $(-4.745,-4.387]$ & Confined & 26 & CME-no-DHII & 42 & 0.35 & 0.34 & 0.30 & $3.0\times10^{-2}$ \\
& $(-4.745,-4.387]$ & Confined & 26 & CME+DHII & 19 & 0.54 & 0.41 & 0.37 & $2.0\times10^{-3}$ \\
& $(-4.745,-4.387]$ & CME+DHII & 19 & CME-no-DHII & 42 & 0.42 & 0.38 & 0.34 & $1.0\times10^{-2}$ \\
& $(-4.387,-2.824]$ & Confined & 10 & CME-no-DHII & 40 & 0.25 & 0.48 & 0.43 & $6.6\times10^{-1}$ \\
& $(-4.387,-2.824]$ & Confined & 10 & CME+DHII & 41 & 0.73 & 0.48 & 0.43 & $1.1\times10^{-4}$ \\
& $(-4.387,-2.824]$ & CME+DHII & 41 & CME-no-DHII & 40 & 0.56 & 0.30 & 0.27 & $1.2\times10^{-6}$ \\
& $(-4.387,-2.824]$ & SEP & 26 & CME-no-DHII & 40 & 0.68 & 0.34 & 0.31 & $1.7\times10^{-7}$ \\
\hline
\multirow{10}{*}{\Fri}
& $(-4.959,-4.745]$ & Confined & 37 & CME-no-DHII & 42 & 0.32 & 0.31 & 0.28 & $2.0\times10^{-2}$ \\
& $(-4.959,-4.745]$ & Confined & 37 & CME+DHII & 10 & 0.53 & 0.48 & 0.43 & $1.0\times10^{-2}$ \\
& $(-4.959,-4.745]$ & CME+DHII & 10 & CME-no-DHII & 42 & 0.31 & 0.48 & 0.43 & $3.2\times10^{-1}$ \\
& $(-4.745,-4.387]$ & Confined & 26 & CME-no-DHII & 42 & 0.18 & 0.34 & 0.30 & $6.0\times10^{-1}$ \\
& $(-4.745,-4.387]$ & Confined & 26 & CME+DHII & 19 & 0.55 & 0.41 & 0.37 & $1.0\times10^{-3}$ \\
& $(-4.745,-4.387]$ & CME+DHII & 19 & CME-no-DHII & 42 & 0.41 & 0.38 & 0.34 & $1.0\times10^{-2}$ \\
& $(-4.387,-2.824]$ & Confined & 9 & CME-no-DHII & 40 & 0.42 & 0.50 & 0.45 & $1.1\times10^{-1}$ \\
& $(-4.387,-2.824]$ & Confined & 9 & CME+DHII & 41 & 0.68 & 0.50 & 0.45 & $8.3\times10^{-4}$ \\
& $(-4.387,-2.824]$ & CME+DHII & 41 & CME-no-DHII & 40 & 0.38 & 0.30 & 0.27 & $4.0\times10^{-3}$ \\
& $(-4.387,-2.824]$ & SEP & 26 & CME-no-DHII & 40 & 0.54 & 0.34 & 0.31 & $9.5\times10^{-5}$ \\
\hline
\multirow{10}{*}{\Frp}
& $(-4.959,-4.745]$ & Confined & 38 & CME-no-DHII & 42 & 0.32 & 0.30 & 0.27 & $2.0\times10^{-2}$ \\
& $(-4.959,-4.745]$ & Confined & 38 & CME+DHII & 10 & 0.51 & 0.48 & 0.43 & $2.0\times10^{-2}$ \\
& $(-4.959,-4.745]$ & CME+DHII & 10 & CME-no-DHII & 42 & 0.28 & 0.48 & 0.43 & $4.9\times10^{-1}$ \\
& $(-4.745,-4.387]$ & Confined & 26 & CME-no-DHII & 42 & 0.10 & 0.34 & 0.30 & $9.8\times10^{-1}$ \\
& $(-4.745,-4.387]$ & Confined & 26 & CME+DHII & 19 & 0.48 & 0.41 & 0.37 & $7.6\times10^{-3}$ \\
& $(-4.745,-4.387]$ & CME+DHII & 19 & CME-no-DHII & 42 & 0.41 & 0.38 & 0.34 & $1.7\times10^{-2}$ \\
& $(-4.387,-2.824]$ & Confined & 10 & CME-no-DHII & 40 & 0.43 & 0.48 & 0.43 & $9.0\times10^{-2}$ \\
& $(-4.387,-2.824]$ & Confined & 10 & CME+DHII & 41 & 0.63 & 0.48 & 0.43 & $1.8\times10^{-3}$ \\
& $(-4.387,-2.824]$ & CME+DHII & 41 & CME-no-DHII & 40 & 0.33 & 0.30 & 0.27 & $1.9\times10^{-2}$ \\
& $(-4.387,-2.824]$ & SEP & 26 & CME-no-DHII & 40 & 0.45 & 0.34 & 0.31 & $2.2\times10^{-3}$ \\
\enddata
\vspace{-0.5cm}
\end{deluxetable}

Table~\ref{tab:KS_test} shows the results from the K-S test for each type-III property for every $\log$\,\Fxp\ bin and event pair. 
Along with the K-S D and p values, the critical D parameters are also found for 90 and 95\% confidence levels (\Dcnt\ and \Dcntf). 
For \Bdur, D$\geq$\Dcntf\ for all event pairs except (confined, CME-no-DHII) within every $\log$\,\Fxp\ bin, clearly revealing that the cumulative distribution functions { differ significantly across event groups.}
The same is true for \Fri\ and \Frp\ as well, except for the (CME+DHII, CME-no-DHII) pair in the weakest flare bin. 
For the (confined, CME-no-DHII) pair, though the distinction between distribution functions is typically moderate to weak for various properties within $\log$\,\Fxp\ bins, we find distinctions with 90--95\% confidence in some bins, especially in the weakest flare bin for all parameters and in the stronger bins for \Frp\ and \Bdur.
{ These results and the observation that most flares that lack DH type-III bursts are confined (see Sect.~\ref{sec:results}) strongly suggest that the Sun-as-a-star type-III emission properties can help significantly improve the estimation of the occurrence probabilities of major space weather events during a solar/stellar SXR flare.}

\subsection{Space weather event occurrence probabilities from Sun-as-a-star flare diagnostics}
\label{sec:predictive}
{ Sections~\ref{sec:results} and~\ref{sec:DHIII_props} show that the occurrence and Sun-as-a-star parameters of DH type-III bursts could help improve the assessment of the space weather impacts of SXR flares. Here, using a sample of 615 SXR flares with reliable detection or non-detection of DH type-III bursts, we assess whether a multidimensional flare vector, $\mathbf{f}$, formed using SXR and DH type-III parameters can be used to reliably estimate the occurrence probabilities of different space weather event classes: confined flares, CMEs, and { SEPE}s. 
The components of $\mathbf{f}$ are functions of \Frp, \Fri, \Bdur, \trise, \Fxp, and $g$, where $g$ is a Boolean gate parameter that takes a value of 1 (0) for flares with (without) DH type-III emission. 
The components of $\mathbf{f}$ that depend on type-III properties are multiplied by $g$ to form gated parameters. This gating permits a consistent definition of $\mathbf{f}$ for flares with and without type-III emission alike.
For a space weather event class ${\rm S_i}$ and flare vector $\mathbf{f}$, the occurrence probability, ${\rm P(S_i\mid\mathbf{f})}$, is defined as,
\begin{equation}
{\rm P(S_i\mid\mathbf{\mathbf{f}})}=
 \frac{\exp(\beta_{i,0}+\boldsymbol{\beta}_i^{\rm T}\mathbf{f})}
 {\sum_j\exp(\beta_{j,0}+\boldsymbol{\beta}_j^{\rm T}\mathbf{f})},
 \label{eq:softmax_spaceweather}
\end{equation}
where ${\rm \beta_{i,0}}$ and ${\rm \boldsymbol{\beta}_i}$ are model coefficients. We consider two probability-estimation problems, ${\rm S_i}\in\{\mathrm{confined},\mathrm{CME}\}$ and ${\rm S_i}\in\{\mathrm{confined},\mathrm{CME},\mathrm{SEPE}\}$, and various types of probability models that differ in the components of $\mathbf{f}$ (see Table~\ref{tab:pred_models}). Some models use purely type-III or SXR flare properties, while others use both or physically-motivated scale-free parameters derived from them to define $\mathbf{f}$ and infer ${\rm P(S_i\mid\mathbf{f})}$. For a chosen model type, every flare in the dataset produces an ($\mathbf{f}$, ${\rm S^{obs}}$) tuple, where ${\rm S^{obs}}$ is the observed space weather class given the problem. 
The primary aim of the following analysis is to identify the model type and optimal parameters that provide the most reliable estimates of ${\rm P(S_i\mid\mathbf{f})}$ for the considered problems.

For a given problem and a model type, we use stratified five-fold cross-validation with 20 repetitions, given the relatively small sample size, especially in the SEPE class~\citep[e.g.][]{BROWNE00_CV_ML,Stumpo2021,asensio23_AIML_rev}.
In each repetition, we divide all $\mathbf{f}$ vectors into five folds that approximately preserve the class fractions in the full sample. Four folds are used to optimize the model coefficients using an ${\rm L_2}$-regularized logistic regression algorithm~\citep{hoerl70ridge}, and the optimized model is then used to estimate ${\rm P(}{\rm S_i}\mid\mathbf{f})$ for the flares in the unused test-fold.
This is repeated with each of the five folds serving once as the test-fold, yielding for each $\mathbf{f}$ a set of ${\rm P(S_i\mid\mathbf{f})},\, {\rm i=1,...,K}$, where K is the number of space weather classes in the problem. 
These predicted probabilities and ${\rm S^{obs}}$ for every $\mathbf{f}$ are used to compute different performance scores, detailed in Sect.~\ref{sec:model_perform}.
This five-fold procedure is repeated 20 times after reshuffling the dataset each time. Each repetition gives one value per performance score, generating a set of 20 values for each score. We report the mean and standard deviation of each score.
The whole procedure produces 100 (20$\times$ 5) sets of model coefficients, from which we derive an `mean-coefficient model'. The entire analysis is repeated for every model type and for both probability-estimation problems, resulting in a mean-coefficient model and a set of three mean performance scores with respective estimation errors.

\subsubsection{Performance scores and reliable models for solar and stellar application}\label{sec:model_perform}
The performance scores are evaluated after each five-fold procedure for a probability-estimation problem with K classes and an overall sample of N flare vectors.
Because the primary objective is to estimate occurrence probabilities rather than assign discrete space weather class labels, we evaluate probability accuracy using the Brier score (BS),
\begin{equation}
{\rm BS}=\frac{1}{{\rm N}}\sum_{{\rm n=1}}^{{\rm N}}\sum_{\rm i=1}^{{\rm K}}
\left[{\rm P(S_i\mid\mathbf{f}_n)}-{\rm \delta_{S_i,S_{n}^{obs}}}\right]^2
\label{eq:brier}
\end{equation}
where $\delta$-function is 1 if the i$^{\rm th}$ space weather class, ${\rm S_i = S^{obs}_n}$ and zero otherwise. ${\rm S^{obs}_n}$ denotes the observed class of the $n^{\rm th}$ flare represented by $\mathbf{f}_n$. BS score hence compares a probability model to an ideal probability-estimator case.
Smaller BS indicates more accurate probability estimates. However, ${\rm BS}=0$ is not expected for a stochastic space weather event occurrence problem using Sun-as-a-star parameters that lack spatial information.
Further, we assess the reliability of the estimated event probabilities for each ${\rm S_i}$ class using the calibration error (CE) score. Firstly, for the i$^{\rm th}$ event class, we divide the full set of flare vectors into equal-count bins along the axis of predicted probability values and estimate the absolute difference between the observed event occurrence fraction (${\rm O_{p,i}}$) and the mean probability estimate (${\rm \bar{P}_{p,i}}$) for each bin, labelled by $p$. Combining all such estimates of absolute deviation across bins for every event class, we estimate
\begin{equation}
{\rm CE}=\frac{1}{{\rm K}}\sum_{{\rm i=1}}^{{\rm K}}{\rm \sum_{p=1}^{P}\frac{N_p}{N}\left|\bar P_{p,i}-O_{p,i}\right|},
\label{eq:calerror}
\end{equation}
where N$_{\rm p}$/N measures the fraction of flares in the ${\rm p^{th}}$ bin. The CE score estimates the mean deviation of the probability estimates from the observed event occurrence fractions across various event classes considered in a problem. Lower CE indicates
better calibration. 

As a secondary test, we report the balanced accuracy (BA) score of the class-predictor corresponding to the probability estimation model. The predicted class, $\hat S$ for a given $\mathbf{f}$, is defined as the class with the highest ${\rm P(S_i\mid\mathbf{f})}$. BA is the mean recall across the K event classes,
\begin{equation}
{\rm BA}=\frac{1}{{\rm K}}\sum_{\rm i=1}^{\rm K}
\frac{{\rm TP}_i}{{\rm TP}_i+{\rm FN}_i},
\end{equation}
where TP$_{\rm i}$ and FN$_{\rm i}$ denote the numbers of true positives and false negatives, respectively, for class ${\rm S_i}$. BA is used as a common measure of classification accuracy, but it is secondary here because assigning a single class discards the magnitude and distribution of the inferred probabilities.

\subsubsection{Solar space weather assessment}\label{sec:predictive_solar}
For the confined/CME problem, we use all 615 events with the most reliable SXR and radio data: 379 confined flares and 236 CMEs. But for the confined/CME/SEPE problem, we remove 28 events with \texttt{SEP\_max}=0 in the catalog, indicating uncertain SEPE associations due to overlapping events. The remaining 587 events contain 362 confined flares, 196 CMEs without SEPEs, and 29 { SEPE}s. Each problem hence uses a mutually exclusive set of event classes.
We study three types of models: SXR only, Type-III only, and SXR+Type-III,
which use parameters based on SXR and/or type-III properties as their names
suggest. Table~\ref{tab:pred_models} presents the components of $\mathbf{f}$
for the various models and their performance scores for the two
probability-estimation problems. The mean and standard deviation of each
score over the 20 repeated five-fold trials are reported.

The SXR+Type-III model provides the best overall scores for both problems. Its BS and CE scores are lower than the SXR-only model, given the estimation errors, demonstrating improved event-occurrence probability estimates.
The corresponding event class-predictor gives ${\rm BA}=0.80\pm0.005$ ($0.71\pm0.015$) for the binary (three-class) problem, compared to $0.73\pm0.003$($0.56\pm0.009$) for the SXR-only model. 
Thus, DH type-III diagnostics provide information complementary to SXR flare properties in estimating CME and { SEPE} occurrence probabilities.

\begin{table*}
\centering
\caption{{ Performance metrics of various models for the two occurrence probability estimation problems. BS and CE evaluate the occurrence probabilities directly, while BA evaluates the corresponding event predictor based on the maximum-probability class (see Sect.~\ref{sec:predictive} for details). Lower values of BS and CE, and higher BA indicate better performance.\label{tab:pred_models}}}
\setlength{\tabcolsep}{3pt}
\begin{tabular}{lllcccc}
\hline
Problem & Model type & $\mathbf{\mathrm{f}}$ & $N$ & BS & CE & BA \\
\hline
\multirow{4}{*}{\shortstack{Confined \\ vs. CME}}
& SXR only
& $\{\mathrm{F}_{\rm X,p},\,\mathrm{t}_{\rm rise}\}$
& 615 & $0.16\pm0.001$ & $0.04\pm0.004$ & $0.73\pm0.003$ \\
& Type-III only
& $\{g,\,g\mathrm{F}_{\rm r,p},\,g\mathrm{F}_{\rm r,int},\,g\mathrm{\Delta T}_{\rm burst}\}$
& 615 & $0.15\pm0.001$ & $0.04\pm0.006$ & $0.77\pm0.003$ \\
& \textbf{SXR + Type-III}
& $\{\mathrm{F}_{\rm X,p},\,\mathrm{t}_{\rm rise},\,g,\,g\mathrm{F}_{\rm r,p},\,g\mathrm{F}_{\rm r,int},\,g\mathrm{\Delta T}_{\rm burst}\}$
& 615 & $\mathbf{0.13\pm0.001}$ & $\mathbf{0.03\pm0.004}$ & $\mathbf{0.80\pm0.005}$ \\
& Scale-free
& $\left\{g,\,g\log\!\frac{\mathrm{F}_{\rm r,p}}{\mathrm{F}_{\rm X,p}},\, g\log\!\frac{\mathrm{\Delta T}_{\rm burst}}{\mathrm{t}_{\rm rise}},\,\log\mathrm{R}_{\rm X}\right\}$
& 615 & $0.15\pm0.001$ & $0.03\pm0.004$ & $0.76\pm0.005$ \\
\hline
\multirow{4}{*}{\shortstack{Confined \\ vs. CME \\ vs. SEPE}}
& SXR only
& $\{\mathrm{F}_{\rm X,p},\,\mathrm{t}_{\rm rise}\}$
& 587 & $0.37\pm0.002$ & $0.04\pm0.003$ & $0.56\pm0.009$ \\
& Type-III only
& \{$g,\,g\mathrm{F}_{\rm r,p}$,\,$g\mathrm{F}_{\rm r,int},\,g\mathrm{\Delta T}_{\rm burst}\}$
& 587 & $0.34\pm0.002$ & $0.02\pm0.004$ & $0.67\pm0.008$ \\
& \textbf{SXR + Type-III}
& $\{\mathrm{F}_{\rm X,p},\,\mathrm{t}_{\rm rise},\,g,\,g\mathrm{F}_{\rm r,p},\,g\mathrm{F}_{\rm r,int},\,g\mathrm{\Delta T}_{\rm burst}\}$
& 587 & $\mathbf{0.30\pm0.003}$ & $\mathbf{0.02\pm0.003}$ & $\mathbf{0.71\pm0.015}$ \\
& Scale-free
& $\left\{g,\,g\log\!\frac{\mathrm{F}_{\rm r,p}}{\mathrm{F}_{\rm X,p}},\,g\log\!\frac{\mathrm{\Delta T}_{\rm burst}}{\mathrm{t}_{\rm rise}},\,\log\mathrm{R}_{\rm X}\right\}$
& 587 & $0.35\pm0.002$ & $0.02\pm0.003$ & $0.60\pm0.008$ \\
\hline
\end{tabular}
\end{table*}


\subsubsection{A scale-free framework for solar and exospace weather applications}\label{sec:predictive_exospace}
We now define and evaluate the scale-free model, independent of the solar activity scales, and hence should be extendable across stellar flares with varying physical scales while sharing a phenomenological correspondence to solar events. This model also lets us evaluate the space weather relevance of solar and stellar flares within a common flare vector space framework.
The model defines $\mathbf{f}$ as a combination of parameters that probe the relative physical scales of non-thermal particle acceleration and heating, explored by radio and SXR observables, respectively.
The additional $\mathrm{R}_{\rm X}$ is the ratio of \Fxp\ to the bolometric solar flux at 1\,AU, given its effective surface temperature of 5880\,K. $\mathrm{R}_{\rm X}$ is a common stellar activity metric, known to scale with mean physical parameters, including Rossby number and magnetic field strength~\citep{stepien94_Defn_Rx+Ro_Vs_activity_n_manyCorCurves,Pizzolato03_Prot-Rx_lateDwarfs,Wright18_Prot-Rx_latedMs}. For both probability estimation problems, the scale-free model gives better mean BS, CE, and BA scores than the SXR-only model.
\begin{figure*}[ht!]
\epsscale{1.15}
\plotone{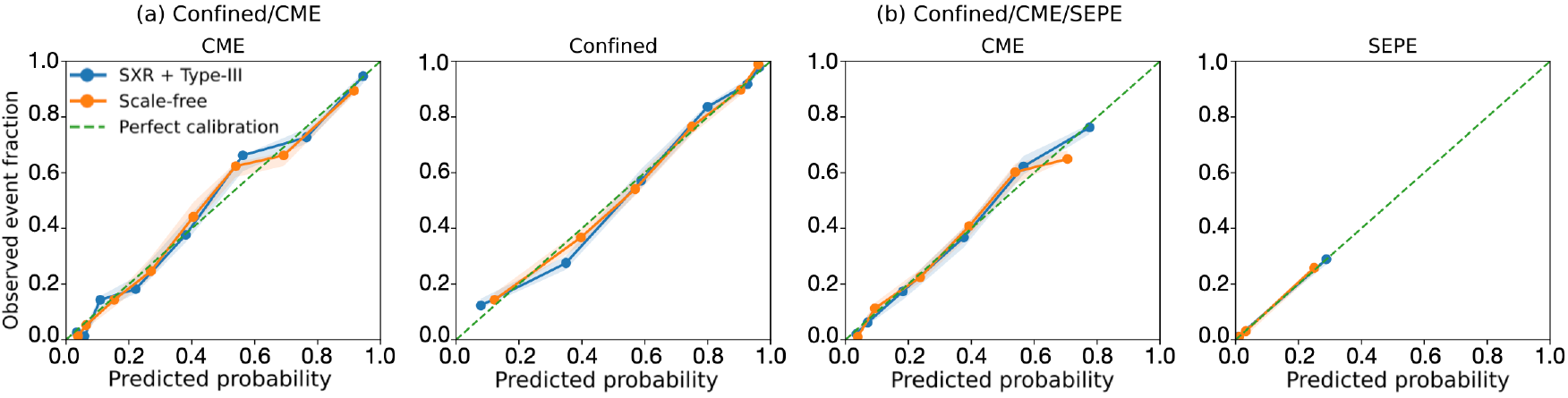}
\caption{{ Calibration curves for mean-coefficient SXR+Type-III and scale-free models for the two probability-estimation problems: (a) confined/CME and (b) confined/CME/{ SEPE}. For each problem and event class, the plotted points compare the predicted occurrence probabilities to the observed event occurrence fraction. Shaded regions show the prediction uncertainties. The dashed diagonal denotes perfect calibration. See Sect.~\ref{sec:predictive_exospace} for details. The legend in panel (a) applies to all panels.}}
\label{fig:prediction_calibration}
\end{figure*}

Further, the model calibration plot in Figure~\ref{fig:prediction_calibration} compares the space weather class probabilities predicted by the mean-coefficient SXR+Type-III and scale-free models with the observed occurrence fractions for each problem. The calibration plot is constructed as follows.
As described in Sect.~\ref{sec:predictive}, each model type and
probability-estimation problem yields 100 sets of optimized coefficients,
whose mean defines the mean-coefficient model.
We apply the mean-coefficient model to the full flare sample and generate a set of predicted probabilities for each ${\rm S_i}$ per flare vector. Then, for each ${\rm S_i}$, the predicted probability values are divided into bins with equal data counts, and the mean predicted probability and
observed event fraction in each bin are plotted. 
To obtain the prediction uncertainty range, we repeat the above procedure using each of the 100 sets of model coefficients, each time generating the corresponding calibration curve.
The shaded uncertainty region shows the 5th-95th percentile range of the 100 resulting calibration curves.
This exercise is performed for each model type, probability-estimation problem, and ${\rm S_i}$. For the binary problem, only the CME class is shown because the confined flare probability and occurrence fraction are complementary to those of the CME class. Both the SXR+Type-III and scale-free models remain close to the perfect calibration line, demonstrating good agreement between the predicted probabilities and observed occurrence fractions.
The mean coefficients and their uncertainties for
the SXR+Type-III and scale-free models are provided in
Appendix~\ref{app:model_coefficients}. The uncertainties ($\sigma$), estimated as the standard deviation of the set of 100 coefficients, reflect the dependence of the inferred coefficients on the data distribution across the training subsets. 

These results demonstrate that a Sun-as-a-star scale-free model based on physically motivated parameters derived from SXR and radio observables retains much of the space weather event occurrence probability information, compared to the scale-dependent solar models.
This provides a proof of concept for a common scale-free Sun-as-a-star framework for solar and exospace weather assessment. 
The present scale-free $\mathbf{f}$ is minimal and needs future work to extend it meaningfully to higher dimensions by defining physics-relevant multiwavelength Sun-as-a-star flare parameters that probe other independent aspects of plasma and magnetic evolution. 
}
\subsection{Properties of S-type-III bursts}
Only 4\% of DH type-II events during 1997–2025 exhibited S-type-III bursts, highlighting their rarity. S-type-III burst events have a median flare class, CME speed, and CME width of M7.9, 1448\,\kms, and 360$^\circ$, respectively. Of the CMEs associated with 26 S-type-III bursts, 81\% (21 out of 26) were halo with deprojected space speeds $>$900\,km\,s$^{-1}$. Space speeds were obtained from the CDAW Halo CME catalog~\citep{Gopal10_HCME_cat}. 
Reliable SEP flux measurements, unaffected by co-temporal events, were available for 22 of the 26 burst events. Of these, 17 (77\%) bursts were associated with major { SEPE}s, with fluxes ranging from 10 to 1480\,pfu.
The 4 out of 5 events without a reported major { SEPE} are east-side events, of which 1 occurred behind the east limb. The only west-side event without a major { SEPE} requires a detailed study, beyond the scope of this work, to infer the contributing factors, including the lack of magnetic connectivity of the IP shock regions to the observing spacecraft.
Our database provides multi-spacecraft radio, SXR, and SEP flux diagnostics. 
Interestingly, 13 out of 14 west-side events produced a major { SEPE}.   
In general, S-type-III bursts are closely associated with very strong flares with fast halo CMEs that can produce major SEPs. 
However, their low occurrence rate limits their utility as a general indicator of major SEPEs and CMEs.

\subsection{Eruptive flares without robust DH type-III features}\label{sec:CME-noDHIII}
\begin{figure*}[!htb]
\epsscale{0.75}
\plotone{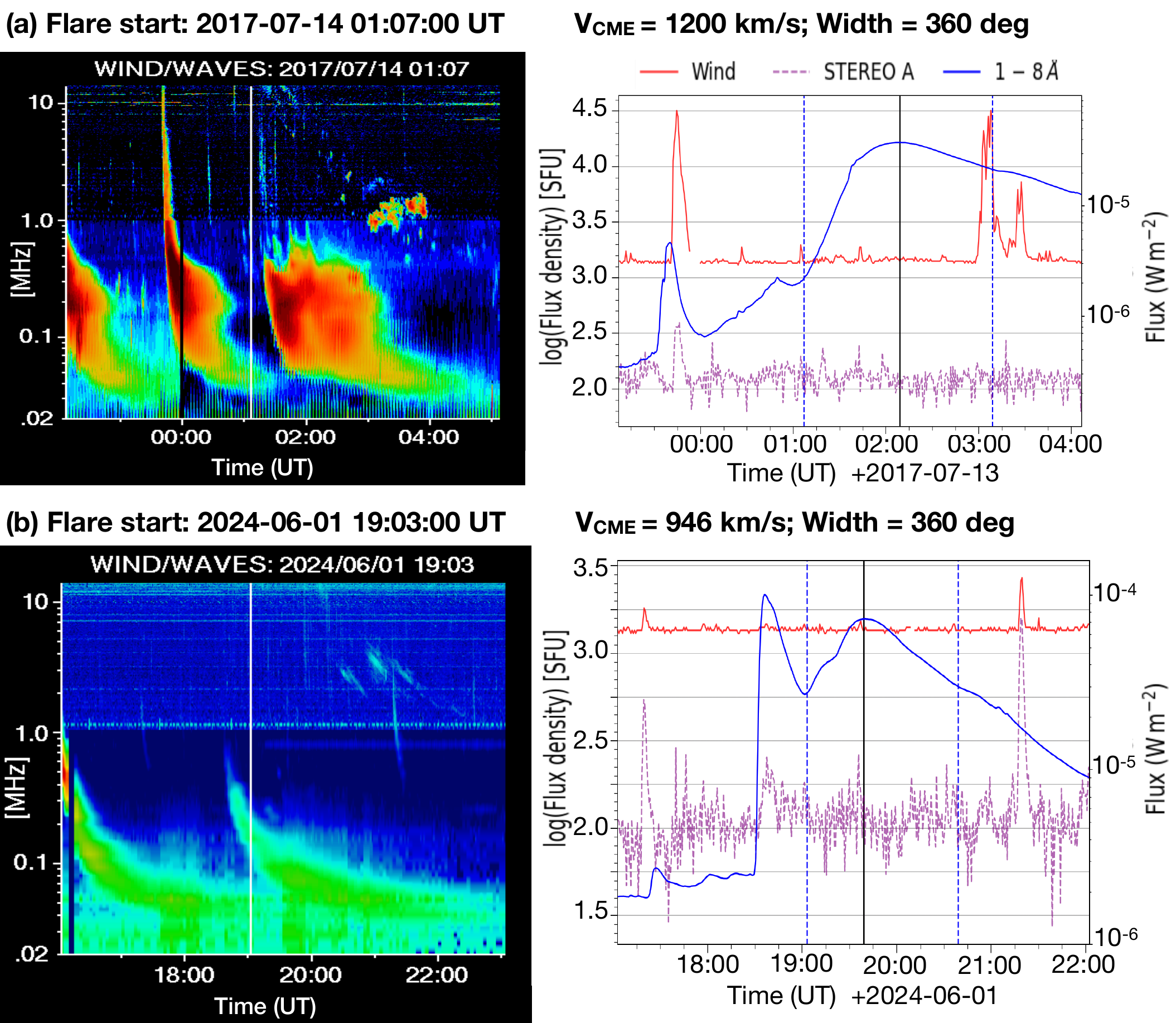}
\vspace{-2mm}
\caption{Examples of halo CME events with DH type-II bursts, but extremely weak or no DH type-III emission. Events with and without an associated major { SEPE} are shown in panels (a) and (b), respectively.
\label{fig_outliers}}
\vspace{-2mm}
\end{figure*}

As mentioned in Sect.~\ref{sec:results}, most (92\%) of the SXR flares without a type-III burst are confined. This can arise from a physical scenario where there are no open field lines connecting the particle acceleration sites to IP space~\citep{2009IAUS..257..283G}.
We find 16 eruptive flares with extremely weak or missing type-III bursts. Though these events constitute only about 5\% (16 out of 311) of the total eruptive flares in our study sample, they are very relevant to stellar CME studies where several radio monitoring studies performed over decades had failed to detect type-II, III, and IV bursts in some very active stars~\citep[e.g.][]{Bastian94_stellarflaresRev,Osten2006,villadsen19_VLA_flarestars}. The type-III bursts being the most common burst type associated with solar CMEs, the physics of type-III-weak CMEs requires detailed case studies.
Some of these events may have metric or kilometric ($<$1\,MHz) type-III bursts without the DH branch due to the details of the magnetic field structure or { opacity along the observing line-of-sight, resulting in reduced radio flux.} 
Detailed case studies are beyond the scope of this paper, but our catalog provides multi-spacecraft, multi-waveband diagnostic data products.
Here, we will just present an overview of DH type-III-weak eruptive flares with some interesting examples.

The DH type-III-weak CMEs have speeds ranging from 163 to 1200 \kms, and associated flare classes range from C5.4 to M7.4.
Three of these CMEs are halo, with two of them associated with DH type-II bursts. 
Figure~\ref{fig_outliers} shows the dynamic spectrum and 
1\,MHz light curves from the STEREO and Wind spacecraft, and the GOES SXR flare light curve for these two CMEs. 
Panel (a) shows the event with the fastest CME and intense DH type-II amongst all eruptive DH type-III weak CMEs. This CME also caused a major { SEPE}. Strong type-III bursts are detected in the kilometric band, though they cannot be robustly traced in the DH band, especially in the 1\,MHz light curve. The bright emission in the 1\,MHz Wind data around 03:00 UT is the DH type-II emission.
Note that the flare source was behind the limb in the field of view of STEREO A. 
Figure~\ref{fig_outliers}b presents the other halo CME with DH type-II bursts but lacking DH type-III emission. This event also shows kilometric type-III bursts, though it is weaker than the previous event. There was no major { SEPE} associated with this CME.
STEREO A and Wind were separated by just $ 14^\circ$, making the dynamic spectra quite similar, and hence it is not ideal for a robust multi-vantage point event analysis.
However, there are several eruptive events in the DH type-III catalog during Cycle 24 when Wind and STEREO spacecraft were well separated to explore the effect of plasma along the line-of-sight and magnetic topology in the observed type-III properties.


\section{Conclusion} \label{sec:conclusion}
We performed an extensive search for DH type-III emission associated with all $\geq$C5-class flares during the period from 2006-12 to 2025-01. Flares with source longitudes within $\pm$55$^\circ$ were only considered. The radio data came from the Wind and STEREO spacecraft. We present a comprehensive catalog of DH type-III bursts and a catalog of flares without robust DH type-III detection. Additionally, we analysed all DH type-II events in the CDAW catalog and compiled a catalog of S-type-III bursts. We provide calibrated DH dynamic spectra, CME height-time plots, links to SXR and white-light movies of the event, and light curves of 1\,MHz flux, SXR flux, and proton flux for every event in these catalogs. The CME (speed, width), SXR flare (peak flux [\Fxp], GOES class, rise time [\trise]), and { SEPE} properties are provided for every event in the catalog.

Of the 734 $\geq$C5 flare events with good-quality SXR and DH data with no confusion from co-temporal emissions, 538 (73\%) show clear co-temporal DH type-III bursts. 196 flares lacked robust DH type-III detections, of which 16 are eruptive. The median DH type-III burst duration (\Bdur), peak flux density (\Frp), integrated flux density (\Fri), and the occurrence time delay of the radio peak flux with respect to SXR (\dtpk) are 18 min, $1.5\times10^5$ SFU, $3.4\times10^7$ SFU\,s, and -0.6\,min, respectively. We analysed the properties of DH type-III bursts within four event groups in the database, namely confined flares (confined), CMEs without DH type-II (CME-no-DHII), CMEs with DH type-II (CME+DHII), and major { SEPE}s (SEP).
Median values and lower envelopes of every DH type-III property, except \dtpk, show a systematic increase moving across confined, CME-no-DHII, CME+DHII, and SEP groups. The lower type-III parameter values for confined flares agree with the fact that only a fraction of the accelerated particles may exit the flare site in these events, while most of the released magnetic energy remains confined. This separation between median \Bdur, \Frp, and \Fri\ values as a function of event group persists even for data within narrow SXR flare-strength bins, indicating that DH type-III burst properties contain information beyond SXR flare strength, which can separate the different event groups. Kolmogorov-Smirnov tests further confirm the statistical distinction of the distributions of these properties, especially \Bdur, for various event group pairs.

{ Further, we tested whether DH type-III parameters can improve the space weather event assessment beyond SXR parameters, using two occurrence probability-estimation problems:
(1) confined/CME and (2) confined/CME/{ SEPE}. For each problem, we derived three event class occurrence probability models differing in their flare parameters: (1) SXR only, (2) Type-III only, and (3) SXR and type-III (SXR+Type-III). We used 20 repetitions of a stratified five-fold cross-validation procedure to derive optimal models in each case, and evaluated the predicted probabilities using Brier score (BS), calibration error (CE) and balanced accuracy (BA).
The SXR+Type-III model provides the best overall probability
estimates, with ${\rm BS}=0.13\pm0.001$ ($0.30\pm0.003$) and
${\rm CE}=0.03\pm0.004$ ($0.02\pm0.003$) for the binary (three-class)
problem, compared with ${\rm BS}=0.16\pm0.001$ ($0.37\pm0.002$) and
${\rm CE}=0.04\pm0.004$ ($0.04\pm0.003$) for the SXR-only model.
The corresponding class predictors give ${\rm BA}=0.80\pm0.005$
($0.71\pm0.015$), compared with $0.73\pm0.003$ ($0.56\pm0.009$) for
the SXR-only case. The mean-coefficient SXR+Type-III model also closely reproduces the observed space weather event fractions in the full-sample calibration tests.
Together, these results demonstrate that DH type-III diagnostics provide
information complementary to SXR flare properties for estimating CME and
{ SEPE} occurrence probabilities.}

{ The models discussed above are based on Sun-as-a-star parameters that depend on absolute solar flare scales. We further demonstrate that a scale-free Sun-as-a-star model based on DH type-III occurrence, dimensionless ratios of type-III and SXR parameters, and ${\rm R_X}$ outperforms the SXR-only model while remaining well calibrated against the observed space weather event occurrence fractions. This demonstrates that useful event-probability information can be retained without relying solely on absolute solar flare scales. This scale-free model framework can be further improved by incorporating independent plasma-evolution diagnostics from other multiwavelength observables, providing a pathway to extend solar space weather diagnostic tools for exospace weather assessment of stellar flares spanning a wide range of physical scales.  Mean model coefficients and their uncertainties for the SXR+Type-III and scale-free models are provided in Appendix~\ref{app:model_coefficients}.}

We find that 26 of 675 DH type-II events (4\%) exhibit clear S-type-III bursts. Although rare, 21 of these events were associated with fast halo CMEs with speeds $>$900\kms, and 77\% of the
22 events with reliable SEP measurements were associated with major { SEPE}s, making S-type-III bursts highly space weather relevant. Meanwhile, 16 out of 311 eruptive flares do not show clear DH type-III emission. We present a few interesting cases of flares with fast halo CMEs but insignificant DH type-III emission, though the type-III features become detectable in the kilometric band.
{ Together with existing multi-spacecraft DH type-II and type-IV catalogs~\citep{Gopal19_DHtypeII_catalog,Atul24_DHtypeIVcat,Atul24_DHtypeIIcat}, the catalogs and statistical framework presented here provide a benchmark for interpreting future DH observations in the context of solar and stellar space weather.}

\begin{acknowledgments}
We thank the CDAW team for maintaining an up-to-date catalog of flares and space weather events. We thank NASA's SPDF center for providing science-ready DH band data from Wind and STEREO spacecraft. 
AM and NG are supported in part by NASA's STEREO project and LWS program.
SA was partially supported by an NSF grant, AGS-2043131.
AM acknowledges the developers of the various Python modules listed in the Software section.
AM acknowledges Seiji Yashiro for posting the catalog online and maintaining the CDAW data resources.
\end{acknowledgments}

\begin{contribution}
AM is responsible for compiling all the catalogs presented here, identifying the DH bursts in the Wind and STEREO databases, doing the analysis, and writing the manuscript. The initial master flare catalog was generated by SA and used by AM to search for radio bursts. SA also cross-matched flare lists with the LASCO-CME catalog. PM is responsible for maintaining the DH type-II, SEP, and GLE event lists. PM also assisted in developing the introduction. NG helped with scientific discussions that improved the manuscript. JB was a short-term intern who briefly assisted with the visual identification of S-type-III bursts.



\end{contribution}

%
\facilities{Wind (WAVES), STEREO (SWAVES), SOHO (LASCO), SDO (AIA)}

\software{NumPy~\citep{numpy}, Astropy~\citep{astropy}, Matplotlib~\citep{matplotlib}, Scikit-learn~\citep{pedregosa11_scikit}, Multiprocessing~\citep{McKerns12_multiprocessing}, Pandas~\citep{reback20_pandas},
Urllib\footnote{\href{https://docs.python.org/3/library/urllib.html}{https://docs.python.org/3/library/urllib.html}}, Beautiful Soup\footnote{\href{https://www.crummy.com/software/BeautifulSoup/bs4/doc/}{https://www.crummy.com/software/BeautifulSoup/bs4/doc/}},
SunPy~\citep{sunpy2020},
Datetime\footnote{\href{https://pypi.org/project/DateTime/}{https://pypi.org/project/DateTime/}},
SciPy~\citep{Scipy2020}
          }

\bibliography{allref}{}

\begin{thebibliography}{}
\expandafter\ifx\csname natexlab\endcsname\relax\def\natexlab#1{#1}\fi
\providecommand{\url}[1]{\href{#1}{#1}}
\providecommand{\dodoi}[1]{doi:~\href{http://doi.org/#1}{\nolinkurl{#1}}}
\providecommand{\doeprint}[1]{\href{http://ascl.net/#1}{\nolinkurl{http://ascl.net/#1}}}
\providecommand{\doarXiv}[1]{\href{https://arxiv.org/abs/#1}{\nolinkurl{https://arxiv.org/abs/#1}}}

\bibitem[{C.~E. {Alissandrakis} {et~al.}(2015){Alissandrakis}, {Nindos}, {Patsourakos}, {Kontogeorgos}, \& {Tsitsipis}}]{alissa2015}
{Alissandrakis}, C.~E., {Nindos}, A., {Patsourakos}, S., {Kontogeorgos}, A., \& {Tsitsipis}, P. 2015, \bibinfo{title}{{A tiny event producing an interplanetary type III burst},} \aap, 582, A52, \dodoi{10.1051/0004-6361/201526265}

\bibitem[{M.~J. {Aschwanden} \& S.~L. {Freeland}(2012){Aschwanden} \& {Freeland}}]{ash2012_flarestats}
{Aschwanden}, M.~J., \& {Freeland}, S.~L. 2012, \bibinfo{title}{{Automated Solar Flare Statistics in Soft X-Rays over 37 Years of GOES Observations: The Invariance of Self-organized Criticality during Three Solar Cycles},} \apj, 754, 112, \dodoi{10.1088/0004-637X/754/2/112}

\bibitem[{M.~J. {Aschwanden} {et~al.}(2014){Aschwanden}, {Xu}, \& {Jing}}]{ash14_flareEnergetics_MagE}
{Aschwanden}, M.~J., {Xu}, Y., \& {Jing}, J. 2014, \bibinfo{title}{{Global Energetics of Solar Flares. I. Magnetic Energies},} \apj, 797, 50, \dodoi{10.1088/0004-637X/797/1/50}

\bibitem[{A. {Asensio Ramos} {et~al.}(2023){Asensio Ramos}, {Cheung}, {Chifu}, \& {Gafeira}}]{asensio23_AIML_rev}
{Asensio Ramos}, A., {Cheung}, M. C.~M., {Chifu}, I., \& {Gafeira}, R. 2023, \bibinfo{title}{{Machine learning in solar physics},} Living Reviews in Solar Physics, 20, 4, \dodoi{10.1007/s41116-023-00038-x}

\bibitem[{ {Astropy Collaboration} {et~al.}(2013){Astropy Collaboration}, {Robitaille}, {Tollerud}, {Greenfield}, {Droettboom}, {Bray}, {Aldcroft}, {Davis}, {Ginsburg}, {Price-Whelan}, {Kerzendorf}, {Conley}, {Crighton}, {Barbary}, {Muna}, {Ferguson}, {Grollier}, {Parikh}, {Nair}, {Unther}, {Deil}, {Woillez}, {Conseil}, {Kramer}, {Turner}, {Singer}, {Fox}, {Weaver}, {Zabalza}, {Edwards}, {Azalee Bostroem}, {Burke}, {Casey}, {Crawford}, {Dencheva}, {Ely}, {Jenness}, {Labrie}, {Lim}, {Pierfederici}, {Pontzen}, {Ptak}, {Refsdal}, {Servillat}, \& {Streicher}}]{astropy}
{Astropy Collaboration}, {Robitaille}, T.~P., {Tollerud}, E.~J., {et~al.} 2013, \bibinfo{title}{{Astropy: A community Python package for astronomy},} \aap, 558, A33, \dodoi{10.1051/0004-6361/201322068}

\bibitem[{M.~G. {Aubier} {et~al.}(1978){Aubier}, {Leblanc}, \& {Moller-Pedersen}}]{Aubier78_typeIIIstorm-typeIlink}
{Aubier}, M.~G., {Leblanc}, Y., \& {Moller-Pedersen}, B. 1978, \bibinfo{title}{{Type I and Type Ill Storm Radiation},} \aap, 70, 685

\bibitem[{T.~S. {Bastian}(1994){Bastian}}]{Bastian94_stellarflaresRev}
{Bastian}, T.~S. 1994, \bibinfo{title}{{Stellar flares},} \ssr, 68, 261, \dodoi{10.1007/BF00749152}

\bibitem[{M.~J. {Bentum} {et~al.}(2020){Bentum}, {Verma}, {Rajan}, {Boonstra}, {Verhoeven}, {Gill}, {van der Veen}, {Falcke}, {Wolt}, {Monna}, {Engelen}, {Rotteveel}, \& {Gurvits}}]{bentum20_OLFAR}
{Bentum}, M.~J., {Verma}, M.~K., {Rajan}, R.~T., {et~al.} 2020, \bibinfo{title}{{A roadmap towards a space-based radio telescope for ultra-low frequency radio astronomy},} Advances in Space Research, 65, 856, \dodoi{10.1016/j.asr.2019.09.007}

\bibitem[{A.~O. {Benz} {et~al.}(2007){Benz}, {Braj{\v{s}}a}, \& {Magdaleni{\'c}}}]{benz07-radioqietflares}
{Benz}, A.~O., {Braj{\v{s}}a}, R., \& {Magdaleni{\'c}}, J. 2007, \bibinfo{title}{{Are There Radio-quiet Solar Flares?},} \solphys, 240, 263, \dodoi{10.1007/s11207-007-0365-9}

\bibitem[{J.~L. {Bougeret} {et~al.}(1995){Bougeret}, {Kaiser}, {Kellogg}, {Manning}, {Goetz}, {Monson}, {Monge}, {Friel}, {Meetre}, {Perche}, {Sitruk}, \& {Hoang}}]{WAVES1995}
{Bougeret}, J.~L., {Kaiser}, M.~L., {Kellogg}, P.~J., {et~al.} 1995, \bibinfo{title}{{Waves: The Radio and Plasma Wave Investigation on the Wind Spacecraft},} \ssr, 71, 231, \dodoi{10.1007/BF00751331}

\bibitem[{M.~W. Browne(2000)Browne}]{BROWNE00_CV_ML}
Browne, M.~W. 2000, \bibinfo{title}{Cross-Validation Methods,} Journal of Mathematical Psychology, 44, 108, \dodoi{https://doi.org/10.1006/jmps.1999.1279}

\bibitem[{I.~H. {Cairns} \& D.~B. {Melrose}(1985){Cairns} \& {Melrose}}]{cairns85_secondHarmplasmaemiss}
{Cairns}, I.~H., \& {Melrose}, D.~B. 1985, \bibinfo{title}{{A theory for the 2f$_{p}$ radiation upstream of the earth's bow shock},} \jgr, 90, 6637, \dodoi{10.1029/JA090iA07p06637}

\bibitem[{H.~V. {Cane} {et~al.}(2002){Cane}, {Erickson}, \& {Prestage}}]{Cane02_flare-typeIII-CME-SEP-link}
{Cane}, H.~V., {Erickson}, W.~C., \& {Prestage}, N.~P. 2002, \bibinfo{title}{{Solar flares, type III radio bursts, coronal mass ejections, and energetic particles},} Journal of Geophysical Research (Space Physics), 107, 1315, \dodoi{10.1029/2001JA000320}

\bibitem[{H.~V. {Cane} {et~al.}(1981){Cane}, {Stone}, {Fainberg}, {Stewart}, {Steinberg}, \& {Hoang}}]{cane81_herringBtypeII-typeIII-SEPlink}
{Cane}, H.~V., {Stone}, R.~G., {Fainberg}, J., {et~al.} 1981, \bibinfo{title}{{Radio evidence for shock acceleration of electrons in the solar corona},} \grl, 8, 1285, \dodoi{10.1029/GL008i012p01285}

\bibitem[{C. {Cattell} {et~al.}(2021){Cattell}, {Glesener}, {Leiran}, {Dombeck}, {Goetz}, {Mart{\'\i}nez Oliveros}, {Badman}, {Pulupa}, \& {Bale}}]{Cattel21_typeIII-noSXR}
{Cattell}, C., {Glesener}, L., {Leiran}, B., {et~al.} 2021, \bibinfo{title}{{Periodicities in an active region correlated with Type III radio bursts observed by Parker Solar Probe},} \aap, 650, A6, \dodoi{10.1051/0004-6361/202039510}

\bibitem[{C. {Chen} {et~al.}(2025){Chen}, {Su}, {Chen}, {Li}, {Yu}, \& {Gan}}]{Chen25_C-flares-typeIII-HXR-noCME}
{Chen}, C., {Su}, Y., {Chen}, W., {et~al.} 2025, \bibinfo{title}{{Intense Hard X-Ray Emissions in C-class Flares: A Statistical Study with ASO-S/HXI Data},} \apjl, 987, L4, \dodoi{10.3847/2041-8213/addf2f}

\bibitem[{E.~W. Cliver \& A.~G. Ling(2008)Cliver \& Ling}]{Cliver09}
Cliver, E.~W., \& Ling, A.~G. 2008, \bibinfo{title}{LOW-FREQUENCY TYPE III BURSTS AND SOLAR ENERGETIC PARTICLE EVENTS,} 690, 598, \dodoi{10.1088/0004-637X/690/1/598}

\bibitem[{D.~R. {DeBoer} {et~al.}(2025){DeBoer}, {Warnick}, {Tremblay}, {Johnson}, {Reenock}, \& {Keane}}]{Deboer25_FARSIDE}
{DeBoer}, D.~R., {Warnick}, K.~F., {Tremblay}, C.~D., {et~al.} 2025, \bibinfo{title}{{The Lunar Farside Technosignature and Transient Telescope (LFT3)},} in 2025 United States National Committee of URSI National Radio Science Meeting (USNC-URSI NRSM) (IEEE), 46, \dodoi{10.23919/USNC-URSINRSM66067.2025.10907042}

\bibitem[{G. {Del Zanna} {et~al.}(2011){Del Zanna}, {Aulanier}, {Klein}, \& {T{\"o}r{\"o}k}}]{delzanna11_unifiedMod-typeI-typeIIIstrm}
{Del Zanna}, G., {Aulanier}, G., {Klein}, K.-L., \& {T{\"o}r{\"o}k}, T. 2011, \bibinfo{title}{{A single picture for solar coronal outflows and radio noise storms},} \aap, 526, A137, \dodoi{10.1051/0004-6361/201015231}

\bibitem[{J. {Fainberg} \& R.~G. {Stone}(1970){Fainberg} \& {Stone}}]{fainberg70_typeIIIstorm}
{Fainberg}, J., \& {Stone}, R.~G. 1970, \bibinfo{title}{{Type III Solar Radio Burst Storms Observed at Low Frequencies.},} \solphys, 15, 222, \dodoi{10.1007/BF00149487}

\bibitem[{V.~L. {Ginzburg} \& V.~V. {Zhelezniakov}(1958){Ginzburg} \& {Zhelezniakov}}]{ginzburg1958}
{Ginzburg}, V.~L., \& {Zhelezniakov}, V.~V. 1958, \bibinfo{title}{{On the Possible Mechanisms of Sporadic Solar Radio Emission (Radiation in an Isotropic Plasma)},} \sovast, 2, 653

\bibitem[{N. {Gopalswamy}(2004){Gopalswamy}}]{gopal04_IPbursts_overview}
{Gopalswamy}, N. 2004, \bibinfo{title}{{Interplanetary Radio Bursts},} in Astrophysics and Space Science Library, Vol. 314, Astrophysics and Space Science Library, ed. D.~E. {Gary} \& C.~U. {Keller}, 305, \dodoi{10.1007/1-4020-2814-8_15}

\bibitem[{N. {Gopalswamy} {et~al.}(2005){Gopalswamy}, {Aguilar-Rodriguez}, {Yashiro}, {Nunes}, {Kaiser}, \& {Howard}}]{gopal05_m-kmTypeII-EnergeticLink}
{Gopalswamy}, N., {Aguilar-Rodriguez}, E., {Yashiro}, S., {et~al.} 2005, \bibinfo{title}{{Type II radio bursts and energetic solar eruptions},} Journal of Geophysical Research (Space Physics), 110, A12S07, \dodoi{10.1029/2005JA011158}

\bibitem[{N. {Gopalswamy} {et~al.}(2009{\natexlab{a}}){Gopalswamy}, {Akiyama}, \& {Yashiro}}]{2009IAUS..257..283G}
{Gopalswamy}, N., {Akiyama}, S., \& {Yashiro}, S. 2009{\natexlab{a}}, \bibinfo{title}{{Major solar flares without coronal mass ejections},} in IAU Symposium, Vol. 257, Universal Heliophysical Processes, ed. N.~{Gopalswamy} \& D.~F. {Webb}, 283--286, \dodoi{10.1017/S174392130902941X}

\bibitem[{N. {Gopalswamy} {et~al.}(2023){Gopalswamy}, {Kumari}, \& {M{\"a}kel{\"a}}}]{gopal23_sgre-gle-ctypeIII}
{Gopalswamy}, N., {Kumari}, A., \& {M{\"a}kel{\"a}}, P.~A. 2023, \bibinfo{title}{{Type III Radio Bursts from Solar Eruptions and their Connection to GLE and SGRE Events},} in 2023 XXXVth General Assembly and Scientific Symposium of the International Union of Radio Science (URSI GASS), 79, \dodoi{10.23919/URSIGASS57860.2023.10265368}

\bibitem[{N. {Gopalswamy} {et~al.}(2015){Gopalswamy}, {M{\"a}kel{\"a}}, {Akiyama}, {Yashiro}, {Xie}, {Thakur}, \& {Kahler}}]{2015ApJ...806....8G}
{Gopalswamy}, N., {M{\"a}kel{\"a}}, P., {Akiyama}, S., {et~al.} 2015, \bibinfo{title}{{Large Solar Energetic Particle Events Associated with Filament Eruptions Outside of Active Regions},} \apj, 806, 8, \dodoi{10.1088/0004-637X/806/1/8}

\bibitem[{N. {Gopalswamy} {et~al.}(2019){Gopalswamy}, {M{\"a}kel{\"a}}, \& {Yashiro}}]{Gopal19_DHtypeII_catalog}
{Gopalswamy}, N., {M{\"a}kel{\"a}}, P., \& {Yashiro}, S. 2019, \bibinfo{title}{{A Catalog of Type II radio bursts observed by Wind/WAVES and their Statistical Properties},} Sun and Geosphere, 14, 111, \dodoi{10.31401/SunGeo.2019.02.03}

\bibitem[{N. {Gopalswamy} {et~al.}(2026){Gopalswamy}, {Micha{\l}ek}, {Yashiro}, {M{\"a}kel{\"a}}, {Akiyama}, {Xie}, \& {Vourlidas}}]{gopal26_LASCOCMEcat_ver2}
{Gopalswamy}, N., {Micha{\l}ek}, G., {Yashiro}, S., {et~al.} 2026, \bibinfo{title}{{The SOHO LASCO CME Catalog - Version 2},} in IAU Symposium, Vol.~20, IAU Symposium, ed. N.~{Gopalsway}, O.~{Malandraki}, A.~{Vidotto}, \& W.~{Manchester}, IV., 133--139, \dodoi{10.1017/S1743921325000559}

\bibitem[{N. {Gopalswamy} {et~al.}(2007){Gopalswamy}, {Yashiro}, \& {Akiyama}}]{gopal07_geoeffectiveness_HaloCMEs}
{Gopalswamy}, N., {Yashiro}, S., \& {Akiyama}, S. 2007, \bibinfo{title}{{Geoeffectiveness of halo coronal mass ejections},} Journal of Geophysical Research (Space Physics), 112, A06112, \dodoi{10.1029/2006JA012149}

\bibitem[{N. {Gopalswamy} {et~al.}(2001){Gopalswamy}, {Yashiro}, {Kaiser}, {Howard}, \& {Bougeret}}]{Gopal01_DHII-CMEprops}
{Gopalswamy}, N., {Yashiro}, S., {Kaiser}, M.~L., {Howard}, R.~A., \& {Bougeret}, J.-L. 2001, \bibinfo{title}{{Characteristics of coronal mass ejections associated with long-wavelength type II radio bursts},} \jgr, 106, 29219, \dodoi{10.1029/2001JA000234}

\bibitem[{N. {Gopalswamy} {et~al.}(2009{\natexlab{b}}){Gopalswamy}, {Yashiro}, {Michalek}, {Stenborg}, {Vourlidas}, {Freeland}, \& {Howard}}]{Gopal09_LASCO_CMEcatalog}
{Gopalswamy}, N., {Yashiro}, S., {Michalek}, G., {et~al.} 2009{\natexlab{b}}, \bibinfo{title}{{The SOHO/LASCO CME Catalog},} Earth Moon and Planets, 104, 295, \dodoi{10.1007/s11038-008-9282-7}

\bibitem[{N. {Gopalswamy} {et~al.}(2010){Gopalswamy}, {Yashiro}, {Michalek}, {Xie}, {M{\"a}kel{\"a}}, {Vourlidas}, \& {Howard}}]{Gopal10_HCME_cat}
{Gopalswamy}, N., {Yashiro}, S., {Michalek}, G., {et~al.} 2010, \bibinfo{title}{{A Catalog of Halo Coronal Mass Ejections from SOHO},} Sun and Geosphere, 5, 7

\bibitem[{N. {Gopalswamy} {et~al.}(2008){Gopalswamy}, {Yashiro}, {Xie}, {Akiyama}, {Aguilar-Rodriguez}, {Kaiser}, {Howard}, \& {Bougeret}}]{Gopal08_FWCMEs-noDHII}
{Gopalswamy}, N., {Yashiro}, S., {Xie}, H., {et~al.} 2008, \bibinfo{title}{{Radio-Quiet Fast and Wide Coronal Mass Ejections},} \apj, 674, 560, \dodoi{10.1086/524765}

\bibitem[{N. Gopalswamy {et~al.}(2000)Gopalswamy, Kaiser, Thompson, Burlaga, Szabo, Lara, Vourlidas, Yashiro, \& Bougeret}]{gopal00_DH-richCMEs-stronger}
Gopalswamy, N., Kaiser, M.~L., Thompson, B.~J., {et~al.} 2000, \bibinfo{title}{Radio-rich solar eruptive events,} \grl, 27, 1427, \dodoi{https://doi.org/10.1029/1999GL003665}

\bibitem[{L. {Harra} {et~al.}(2021){Harra}, {Brooks}, {Bale}, {Mandrini}, {Barczynski}, {Sharma}, {Badman}, {Vargas Dom{\'\i}nguez}, \& {Pulupa}}]{Harra21_AR-typeIIIstorm}
{Harra}, L., {Brooks}, D.~H., {Bale}, S.~D., {et~al.} 2021, \bibinfo{title}{{The active region source of a type III radio storm observed by Parker Solar Probe during encounter 2},} \aap, 650, A7, \dodoi{10.1051/0004-6361/202039514}

\bibitem[{C.~R. Harris {et~al.}(2020)Harris, Millman, van~der Walt, Gommers, Virtanen, Cournapeau, Wieser, Taylor, Berg, Smith, Kern, Picus, Hoyer, van Kerkwijk, Brett, Haldane, del R{'{\i}}o, Wiebe, Peterson, G{'{e}}rard-Marchant, Sheppard, Reddy, Weckesser, Abbasi, Gohlke, \& Oliphant}]{numpy}
Harris, C.~R., Millman, K.~J., van~der Walt, S.~J., {et~al.} 2020, \bibinfo{title}{Array programming with {NumPy},} Nature, 585, 357, \dodoi{10.1038/s41586-020-2649-2}

\bibitem[{S.~L. {Hawley} {et~al.}(2003){Hawley}, {Allred}, {Johns-Krull}, {Fisher}, {Abbett}, {Alekseev}, {Avgoloupis}, {Deustua}, {Gunn}, {Seiradakis}, {Sirk}, \& {Valenti}}]{2003ApJ...597..535H}
{Hawley}, S.~L., {Allred}, J.~C., {Johns-Krull}, C.~M., {et~al.} 2003, \bibinfo{title}{{Multiwavelength Observations of Flares on AD Leonis},} \apj, 597, 535, \dodoi{10.1086/378351}

\bibitem[{A. {Hillaris} {et~al.}(2016){Hillaris}, {Bouratzis}, \& {Nindos}}]{Hillaris16_typeIV}
{Hillaris}, A., {Bouratzis}, C., \& {Nindos}, A. 2016, \bibinfo{title}{{Interplanetary Type IV Bursts},} \solphys, 291, 2049, \dodoi{10.1007/s11207-016-0946-6}

\bibitem[{T. {Hirayama}(1974){Hirayama}}]{hirayam74_flarepromerupt_model_n_geom}
{Hirayama}, T. 1974, \bibinfo{title}{{Theoretical Model of Flares and Prominences. I: Evaporating Flare Model},} \solphys, 34, 323, \dodoi{10.1007/BF00153671}

\bibitem[{A.~E. Hoerl \& R.~W. Kennard(1970)Hoerl \& Kennard}]{hoerl70ridge}
Hoerl, A.~E., \& Kennard, R.~W. 1970, \bibinfo{title}{Ridge Regression: Biased Estimation for Nonorthogonal Problems,} Technometrics, 12, 55, \dodoi{10.1080/00401706.1970.10488634}

\bibitem[{R.~A. {Howard} {et~al.}(2008){Howard}, {Moses}, {Vourlidas}, {Newmark}, {Socker}, {Plunkett}, {Korendyke}, {Cook}, {Hurley}, {Davila}, {Thompson}, {St Cyr}, {Mentzell}, {Mehalick}, {Lemen}, {Wuelser}, {Duncan}, {Tarbell}, {Wolfson}, {Moore}, {Harrison}, {Waltham}, {Lang}, {Davis}, {Eyles}, {Mapson-Menard}, {Simnett}, {Halain}, {Defise}, {Mazy}, {Rochus}, {Mercier}, {Ravet}, {Delmotte}, {Auchere}, {Delaboudiniere}, {Bothmer}, {Deutsch}, {Wang}, {Rich}, {Cooper}, {Stephens}, {Maahs}, {Baugh}, {McMullin}, \& {Carter}}]{SECCHI2008}
{Howard}, R.~A., {Moses}, J.~D., {Vourlidas}, A., {et~al.} 2008, \bibinfo{title}{{Sun Earth Connection Coronal and Heliospheric Investigation (SECCHI)},} \ssr, 136, 67, \dodoi{10.1007/s11214-008-9341-4}

\bibitem[{J.~D. Hunter(2007)Hunter}]{matplotlib}
Hunter, J.~D. 2007, \bibinfo{title}{Matplotlib: A 2D graphics environment,} Computing in science \& engineering, 9, 90

\bibitem[{S.~W. {Kahler}(1982){Kahler}}]{kahler82_BFS}
{Kahler}, S.~W. 1982, \bibinfo{title}{{The role of the big flare syndrome in correlations of solar energetic proton fluxes and associated microwave burst parameters},} \jgr, 87, 3439, \dodoi{10.1029/JA087iA05p03439}

\bibitem[{S.~W. {Kahler} {et~al.}(1986){Kahler}, {Cliver}, \& {Cane}}]{1986AdSpR...6f.319K}
{Kahler}, S.~W., {Cliver}, E.~W., \& {Cane}, H.~V. 1986, \bibinfo{title}{{The relationship of shock-associated kilometric radio emission with metric type II bursts and energetic particles},} Advances in Space Research, 6, 319, \dodoi{10.1016/0273-1177(86)90168-7}

\bibitem[{M. Knapp {et~al.}(2024)Knapp, Paritsky, Kononov, \& Kao}]{Knapp2024GOLOW}
Knapp, M., Paritsky, L., Kononov, E., \& Kao, M.~M. 2024, \bibinfo{title}{Great Observatory for Long Wavelengths (GO-LoW): Phase I NIAC Final Report,} \doarXiv{2404.08432}

\bibitem[{T. Li {et~al.}(2019)Li, Liu, Hou, \& Zhang}]{Li19_types_confflares}
Li, T., Liu, L., Hou, Y., \& Zhang, J. 2019, \bibinfo{title}{Two Types of Confined Solar Flares,} \apj, 881, 151, \dodoi{10.3847/1538-4357/ab3121}

\bibitem[{R.~P. {Lin}(1974){Lin}}]{Lin74_SEErev-typIIIassoc}
{Lin}, R.~P. 1974, \bibinfo{title}{{Non-relativistic Solar Electrons},} \ssr, 16, 189, \dodoi{10.1007/BF00240886}

\bibitem[{R.~J. MacDowall {et~al.}(2003)MacDowall, Lara, Manoharan, Nitta, Rosas, \& Bougeret}]{macdowall03_typeIIIl-SEP_link}
MacDowall, R.~J., Lara, A., Manoharan, P.~K., {et~al.} 2003, \bibinfo{title}{Long-duration hectometric type III radio bursts and their association with solar energetic particle (SEP) events,} Geophysical Research Letters, 30, \dodoi{https://doi.org/10.1029/2002GL016624}

\bibitem[{M.~M. {McKerns} {et~al.}(2012){McKerns}, {Strand}, {Sullivan}, {Fang}, \& {Aivazis}}]{McKerns12_multiprocessing}
{McKerns}, M.~M., {Strand}, L., {Sullivan}, T., {Fang}, A., \& {Aivazis}, M. A.~G. 2012, \bibinfo{title}{{Building a Framework for Predictive Science},} arXiv e-prints, arXiv:1202.1056, \dodoi{10.48550/arXiv.1202.1056}

\bibitem[{D. McLean \& N. Labrum(1985)McLean \& Labrum}]{Mclean85_book}
McLean, D., \& Labrum, N. 1985, {Solar Radio Astrophysics} (Cambridge University Press)

\bibitem[{D.~B. {Melrose}(1970){Melrose}}]{melrose1970}
{Melrose}, D.~B. 1970, \bibinfo{title}{{On the theory of type II and type III solar radio bursts. II. Alternative model},} Australian Journal of Physics, 23, 885, \dodoi{10.1071/PH700885}

\bibitem[{R. {Miteva} {et~al.}(2017){Miteva}, {Samwel}, \& {Krupar}}]{Miteva17_SEP-radburst_link}
{Miteva}, R., {Samwel}, S.~W., \& {Krupar}, V. 2017, \bibinfo{title}{{Solar energetic particles and radio burst emission},} Journal of Space Weather and Space Climate, 7, A37, \dodoi{10.1051/swsc/2017035}

\bibitem[{R. {Miteva} {et~al.}(2022){Miteva}, {Samwel}, \& {Zabunov}}]{Miteva22_SEEs-typeIII}
{Miteva}, R., {Samwel}, S.~W., \& {Zabunov}, S. 2022, \bibinfo{title}{{Solar Radio Bursts Associated with In Situ Detected Energetic Electrons in Solar Cycles 23 and 24},} Universe, 8, 275, \dodoi{10.3390/universe8050275}

\bibitem[{A. {Mohan} {et~al.}(2024{\natexlab{a}}){Mohan}, {Gopalswamy}, {Kumari}, {Akiyama}, \& {G}}]{Atul24_DHtypeIVcat}
{Mohan}, A., {Gopalswamy}, N., {Kumari}, A., {Akiyama}, S., \& {G}, S. 2024{\natexlab{a}}, \bibinfo{title}{{Interplanetary Type IV Solar Radio Bursts: A Comprehensive Catalog and Statistical Results},} \apj, 971, 86, \dodoi{10.3847/1538-4357/ad5315}

\bibitem[{A. {Mohan} {et~al.}(2026{\natexlab{a}}){Mohan}, {Gopalswamy}, {Mondal}, {Kumari}, \& {Sindhuja}}]{2026IAUS..388..338M}
{Mohan}, A., {Gopalswamy}, N., {Mondal}, S., {Kumari}, A., \& {Sindhuja}, G. 2026{\natexlab{a}}, \bibinfo{title}{{CME-associated type-IV radio bursts: The solar paradigm and the unique case of AD Leo},} in IAU Symposium, Vol.~20, IAU Symposium, ed. N.~{Gopalsway}, O.~{Malandraki}, A.~{Vidotto}, \& W.~{Manchester}, IV., 338--344, \dodoi{10.1017/S1743921324001832}

\bibitem[{A. {Mohan} {et~al.}(2024{\natexlab{b}}){Mohan}, {Gopalswamy}, {Raju}, \& {Akiyama}}]{Atul24_modGBR}
{Mohan}, A., {Gopalswamy}, N., {Raju}, H., \& {Akiyama}, S. 2024{\natexlab{b}}, \bibinfo{title}{{Novel scaling laws to derive spatially resolved flare and CME parameters from sun-as-a-star observables},} \aap, 691, L8, \dodoi{10.1051/0004-6361/202451072}

\bibitem[{A. {Mohan} {et~al.}(2026{\natexlab{b}}){Mohan}, {Gopalswamy}, {Raju}, \& {Akiyama}}]{Atul24_DHtypeIIcat}
{Mohan}, A., {Gopalswamy}, N., {Raju}, H., \& {Akiyama}, S. 2026{\natexlab{b}}, \bibinfo{title}{{A catalog of multi-vantage point observations of type-II bursts: Statistics and correlations},} 20, 345, \dodoi{10.1017/S1743921324001558}

\bibitem[{A. {Mohan} {et~al.}(2019){Mohan}, {Mondal}, {Oberoi}, \& {Lonsdale}}]{Atul19_typeIIIQPP_turb}
{Mohan}, A., {Mondal}, S., {Oberoi}, D., \& {Lonsdale}, C.~J. 2019, \bibinfo{title}{{Evidence for Super-Alfv{\'e}nic Oscillations in Solar Type III Radio Burst Sources},} \apj, 875, 98, \dodoi{10.3847/1538-4357/ab0ae5}

\bibitem[{A. {Mohan} {et~al.}(2024{\natexlab{c}}){Mohan}, {Mondal}, {Wedemeyer}, \& {Gopalswamy}}]{atul24_ADLeotypeIV}
{Mohan}, A., {Mondal}, S., {Wedemeyer}, S., \& {Gopalswamy}, N. 2024{\natexlab{c}}, \bibinfo{title}{{Energetic particle activity in AD Leo: Detection of a solar-like type-IV burst},} \aap, 686, A51, \dodoi{10.1051/0004-6361/202347924}

\bibitem[{R.~A. {Osten} \& T.~S. {Bastian}(2006){Osten} \& {Bastian}}]{Osten2006}
{Osten}, R.~A., \& {Bastian}, T.~S. 2006, \bibinfo{title}{{Wide-Band Spectroscopy of Two Radio Bursts on AD Leonis},} \apj, 637, 1016, \dodoi{10.1086/498410}

\bibitem[{T. pandas~development team(2020)pandas~development team}]{reback20_pandas}
pandas~development team, T. 2020, pandas-dev/pandas: Pandas, latest Zenodo, \dodoi{10.5281/zenodo.3509134}

\bibitem[{F. Pedregosa {et~al.}(2011)Pedregosa, Varoquaux, Gramfort, Michel, Thirion, Grisel, Blondel, Prettenhofer, Weiss, Dubourg, Vanderplas, Passos, Cournapeau, Brucher, Perrot, \& Duchesnay}]{pedregosa11_scikit}
Pedregosa, F., Varoquaux, G., Gramfort, A., {et~al.} 2011, \bibinfo{title}{Scikit-learn: Machine Learning in Python,} Journal of Machine Learning Research, 12, 2825

\bibitem[{N. {Pizzolato} {et~al.}(2003){Pizzolato}, {Maggio}, {Micela}, {Sciortino}, \& {Ventura}}]{Pizzolato03_Prot-Rx_lateDwarfs}
{Pizzolato}, N., {Maggio}, A., {Micela}, G., {Sciortino}, S., \& {Ventura}, P. 2003, \bibinfo{title}{{The stellar activity-rotation relationship revisited: Dependence of saturated and non-saturated X-ray emission regimes on stellar mass for late-type dwarfs},} \aap, 397, 147, \dodoi{10.1051/0004-6361:20021560}

\bibitem[{H.~A.~S. {Reid} \& E.~P. {Kontar}(2017){Reid} \& {Kontar}}]{2017A&A...606A.141R}
{Reid}, H. A.~S., \& {Kontar}, E.~P. 2017, \bibinfo{title}{{Imaging spectroscopy of type U and J solar radio bursts with LOFAR},} \aap, 606, A141, \dodoi{10.1051/0004-6361/201730701}

\bibitem[{H.~A.~S. {Reid} \& H. {Ratcliffe}(2014){Reid} \& {Ratcliffe}}]{Reid2014}
{Reid}, H.~A.~S., \& {Ratcliffe}, H. 2014, \bibinfo{title}{{A review of solar type III radio bursts},} Research in Astronomy and Astrophysics, 14, 773, \dodoi{10.1088/1674-4527/14/7/003}

\bibitem[{M.~J. {Reiner} \& M.~L. {Kaiser}(1999){Reiner} \& {Kaiser}}]{1999GeoRL..26..397R}
{Reiner}, M.~J., \& {Kaiser}, M.~L. 1999, \bibinfo{title}{{Complex Type III-like radio emissions observed from 1 to 14 MHz},} \grl, 26, 397, \dodoi{10.1029/1998GL900317}

\bibitem[{M.~J. {Reiner} {et~al.}(2000){Reiner}, {Karlick{\'y}}, {Ji{\v{r}}i{\v{c}}ka}, {Aurass}, {Mann}, \& {Kaiser}}]{Reiner00_cmplxtypeIII_origin}
{Reiner}, M.~J., {Karlick{\'y}}, M., {Ji{\v{r}}i{\v{c}}ka}, K., {et~al.} 2000, \bibinfo{title}{{On the Solar Origin of Complex Type III-like Radio Bursts Observed at and below 1 MHZ},} \apj, 530, 1049, \dodoi{10.1086/308394}

\bibitem[{L. {Rodr{\'\i}guez-Garc{\'\i}a} {et~al.}(2025){Rodr{\'\i}guez-Garc{\'\i}a}, {G{\'o}mez-Herrero}, {Dresing}, {Balmaceda}, {Palmerio}, {Kouloumvakos}, {Jebaraj}, {Espinosa Lara}, {Roco}, {Palmroos}, {Warmuth}, {Nicolaou}, {Mason}, {Guo}, {Laitinen}, {Cernuda}, {Nieves-Chinchilla}, {Fedeli}, {Lee}, {Cohen}, {Owen}, {Ho}, {Malandraki}, {Vainio}, \& {Rodr{\'\i}guez-Pacheco}}]{rodriguez25_SEP-cmplxtypeIII}
{Rodr{\'\i}guez-Garc{\'\i}a}, L., {G{\'o}mez-Herrero}, R., {Dresing}, N., {et~al.} 2025, \bibinfo{title}{{Solar energetic particles injected inside and outside a magnetic cloud: The widespread solar energetic particle event on 2022 January 20},} \aap, 694, A64, \dodoi{10.1051/0004-6361/202452158}

\bibitem[{P. {Saint-Hilaire} {et~al.}(2013){Saint-Hilaire}, {Vilmer}, \& {Kerdraon}}]{Hilaire}
{Saint-Hilaire}, P., {Vilmer}, N., \& {Kerdraon}, A. 2013, \bibinfo{title}{{A Decade of Solar Type III Radio Bursts Observed by the Nan{\c c}ay Radioheliograph 1998-2008},} \apj, 762, 60, \dodoi{10.1088/0004-637X/762/1/60}

\bibitem[{K. {Shibata} {et~al.}(1995){Shibata}, {Masuda}, {et~al.}}]{shibata95_flarereconn_link}
{Shibata}, K., {Masuda}, S., {et~al.} 1995, \bibinfo{title}{{Hot-Plasma Ejections Associated with Compact-Loop Solar Flares},} \apjl, 451, L83, \dodoi{10.1086/309688}

\bibitem[{K. {Stepien}(1994){Stepien}}]{stepien94_Defn_Rx+Ro_Vs_activity_n_manyCorCurves}
{Stepien}, K. 1994, \bibinfo{title}{{Applicability of the Rossby number in activity-rotation relations for dwarfs and giants.},} \aap, 292, 191

\bibitem[{R.~T. {Stewart} \& N.~R. {Labrum}(1972){Stewart} \& {Labrum}}]{stewart72_typeI-typeIIIstormlink}
{Stewart}, R.~T., \& {Labrum}, N.~R. 1972, \bibinfo{title}{{Meter-wavelength observations of the solar radio burst storm of August 17 22, 1968},} \solphys, 27, 192, \dodoi{10.1007/BF00151783}

\bibitem[{M. Stumpo {et~al.}(2021)Stumpo, Benella, Laurenza, Alberti, Consolini, \& Marcucci}]{Stumpo2021}
Stumpo, M., Benella, S., Laurenza, M., {et~al.} 2021, \bibinfo{title}{Open Issues in Statistical Forecasting of Solar Proton Events: A Machine Learning Perspective,} Space Weather, 19, e2021SW002794, \dodoi{https://doi.org/10.1029/2021SW002794}

\bibitem[{N. {Talebpour Sheshvan} \& S. {Pohjolainen}(2018){Talebpour Sheshvan} \& {Pohjolainen}}]{Nasrin18_DHtypeIV_occult_streamerCMEshock}
{Talebpour Sheshvan}, N., \& {Pohjolainen}, S. 2018, \bibinfo{title}{{Visibility and Origin of Compact Interplanetary Radio Type IV Bursts},} \solphys, 293, 148, \dodoi{10.1007/s11207-018-1371-9}

\bibitem[{ {The SunPy Community} {et~al.}(2020){The SunPy Community}, Barnes, Bobra, Christe, Freij, Hayes, Ireland, Mumford, Perez-Suarez, Ryan, Shih, Chanda, Glogowski, Hewett, Hughitt, Hill, Hiware, Inglis, Kirk, Konge, Mason, Maloney, Murray, Panda, Park, Pereira, Reardon, Savage, Sipőcz, Stansby, Jain, Taylor, Yadav, Rajul, \& Dang}]{sunpy2020}
{The SunPy Community}, Barnes, W.~T., Bobra, M.~G., {et~al.} 2020, \bibinfo{title}{The SunPy Project: Open Source Development and Status of the Version 1.0 Core Package,} The Astrophysical Journal, 890, 68, \dodoi{10.3847/1538-4357/ab4f7a}

\bibitem[{V.~N. {Tsytovich} \& S.~A. {Kaplan}(1969){Tsytovich} \& {Kaplan}}]{Tsytovich69}
{Tsytovich}, V.~N., \& {Kaplan}, S.~A. 1969, \bibinfo{title}{{Conversion of Plasma Waves into Electromagnetic Waves in a Strong Magnetic Field.},} \sovast, 12, 618

\bibitem[{J. {Villadsen} \& G. {Hallinan}(2019){Villadsen} \& {Hallinan}}]{villadsen19_VLA_flarestars}
{Villadsen}, J., \& {Hallinan}, G. 2019, \bibinfo{title}{{Ultra-wideband Detection of 22 Coherent Radio Bursts on M Dwarfs},} ApJ, 871, 214, \dodoi{10.3847/1538-4357/aaf88e}

\bibitem[{P. Virtanen {et~al.}(2020)Virtanen, Gommers, Oliphant, Haberland, Reddy, Cournapeau, Burovski, Peterson, Weckesser, Bright, {van der Walt}, Brett, Wilson, Millman, Mayorov, Nelson, Jones, Kern, Larson, Carey, Polat, Feng, Moore, {VanderPlas}, Laxalde, Perktold, Cimrman, Henriksen, Quintero, Harris, Archibald, Ribeiro, Pedregosa, {van Mulbregt}, \& {SciPy 1.0 Contributors}}]{Scipy2020}
Virtanen, P., Gommers, R., Oliphant, T.~E., {et~al.} 2020, \bibinfo{title}{{{SciPy} 1.0: Fundamental Algorithms for Scientific Computing in Python},} Nature Methods, 17, 261, \dodoi{10.1038/s41592-019-0686-2}

\bibitem[{J.~P. {Wild}(1970){Wild}}]{wild1970}
{Wild}, J.~P. 1970, \bibinfo{title}{{Some Investigations of the Solar Corona: The First Two Years of Observation With the Culgoora Radioheliograph},} Proceedings of the Astronomical Society of Australia, 1, 365, \dodoi{10.1017/S1323358000012364}

\bibitem[{L.~M. {Winter} \& K. {Ledbetter}(2015){Winter} \& {Ledbetter}}]{Winter15_SEP-DHtypeIII-typeII}
{Winter}, L.~M., \& {Ledbetter}, K. 2015, \bibinfo{title}{{Type II and Type III Radio Bursts and their Correlation with Solar Energetic Proton Events},} \apj, 809, 105, \dodoi{10.1088/0004-637X/809/1/105}

\bibitem[{N.~J. {Wright} {et~al.}(2018){Wright}, {Newton}, {Williams}, {Drake}, \& {Yadav}}]{Wright18_Prot-Rx_latedMs}
{Wright}, N.~J., {Newton}, E.~R., {Williams}, P. K.~G., {Drake}, J.~J., \& {Yadav}, R.~K. 2018, \bibinfo{title}{{The stellar rotation-activity relationship in fully convective M dwarfs},} \mnras, 479, 2351, \dodoi{10.1093/mnras/sty1670}

\bibitem[{S. Yashiro {et~al.}(2004)Yashiro, Gopalswamy, Michalek, St.~Cyr, Plunkett, Rich, \& Howard}]{Yashiro04_LASCOCME_Catalog}
Yashiro, S., Gopalswamy, N., Michalek, G., {et~al.} 2004, \bibinfo{title}{A catalog of white light coronal mass ejections observed by the SOHO spacecraft,} \jgr, 109, \dodoi{https://doi.org/10.1029/2003JA010282}

\end{thebibliography}
\bibliographystyle{aasjournalv7}

\appendix
\section{The various event catalogs}\label{app0}
\begin{figure}[htbp]
    \centering
    \epsscale{1.17}
    \plotone{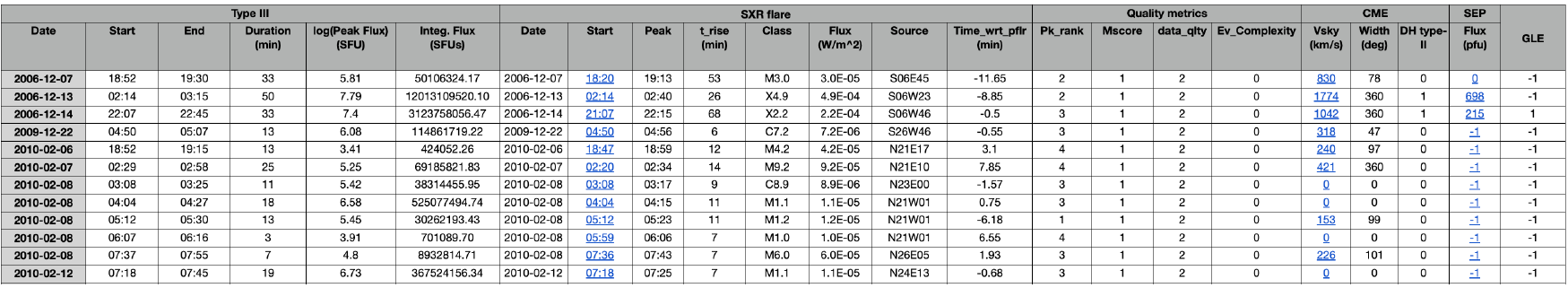}
    \caption{DH type-III catalog. The details of the radio burst, SXR flare, CME, SEP, and GLE are provided for every event.}\label{App1}
\vspace{-3mm}
\end{figure}

\begin{figure}[htbp]
\epsscale{1.17}
\plotone{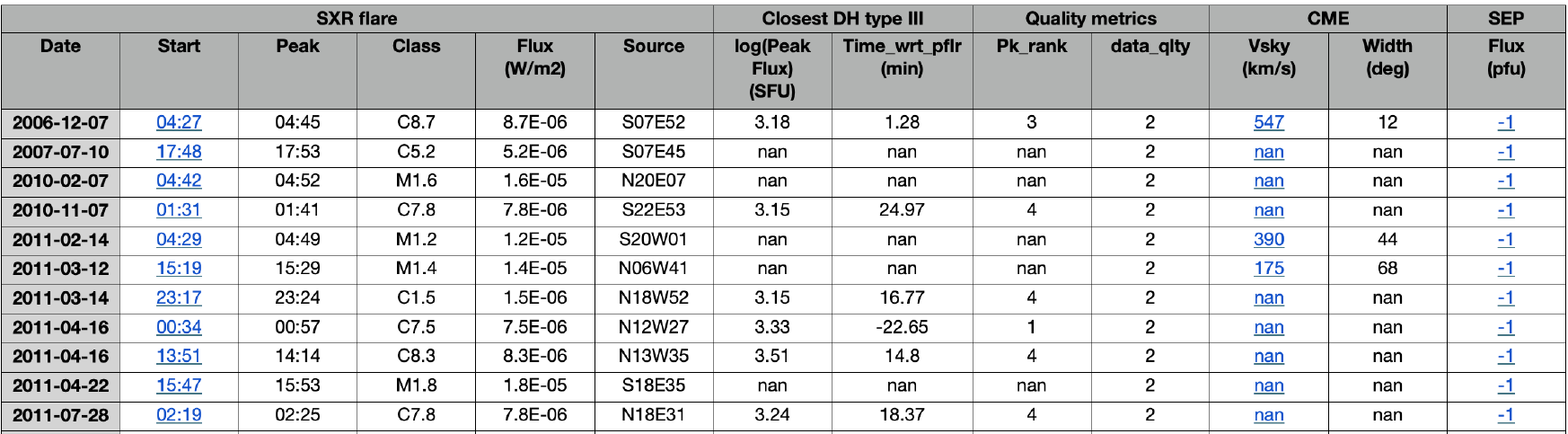}
\vspace{-2mm}
\caption{Catalog of SXR flares with insignificant or no DH type-III. The details of the SXR flare, CME, and SEP are provided for every event. The flux and \dtpk\ of the closest DH type-III burst is also provided for relevant events. Note that Wind/WAVES data usually has a background flux level of log\,F = 3.15 - 3.2\,SFU.}\label{App2}
\vspace{-3mm}
\end{figure}

\begin{figure}[htbp]
\epsscale{1.05}
\plotone{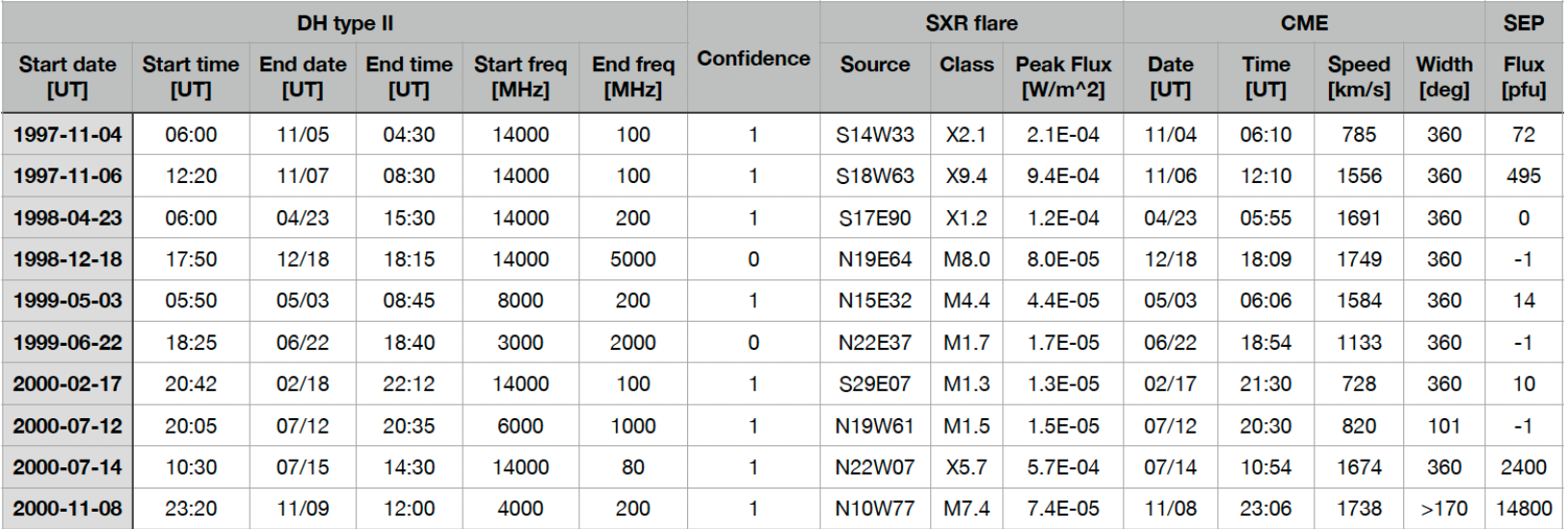}
\vspace{-2mm}
\caption{S-type-III catalog. The details of the DH type-II burst, SXR flare, CME, and SEP are provided for every event.}\label{App3}
\vspace{-3mm}
\end{figure}

This publication presents 3 catalogs. Two of them concern SXR flares with and without DH type-III detections (see Figs.~\ref{App1} and \ref{App2}). Flares with 
source longitudes within $\pm 55^{\circ}$ that occurred between late 2006 and January 2025 were studied. The third catalog, shown in Fig.~\ref{App3}, presents all S-type-III bursts associated with DH type-II events in the CDAW catalog covering the period from 1997 to 2025. 

\section{DH type-III burst properties within flare \Fxp\ bins}\label{app1}
Table~\ref{tab:app1} presents the median DH type-III burst
properties for each event group within individual $\log$\,\Fxp\ bins.

\setlength{\tabcolsep}{4pt}
\renewcommand{\arraystretch}{0.85}
\begin{deluxetable}{llllll}[!ht]
\scriptsize
\tablewidth{0pt}
\tablecaption{Median type-III burst properties in log \Fxp\ bins. Each cell lists the number of events and the median value of each type-III property for a given event group and an \Fxp\ bin. The bin ranges in bold highlight the statistics for $>$M1.1 class flares. Note that the (-5.302, -4.959] bin is used only for the CME+DHII group to have $>$5 data points for median computation.
\label{tab:app1}}
\tablehead{
\colhead{$\log({\rm F_{X,p}})$ bin} &
\colhead{Property} &
\colhead{Confined} &
\colhead{CME-no-DHII} &
\colhead{CME+DHII} &
\colhead{SEP}
}
\startdata
$(-5.302,-5.097]$ & \Bdur\, [min] & $N=72$, Med$=10.0$ & $N=19$, Med$=17.0$ & $N=3$, Med$=24.0$ & $N=0$ \\
 & \Frp\, [SFU] & $N=72$, Med$=2.0\times10^{4}$ & $N=19$, Med$=2.8\times10^{5}$ & $N=3$, Med$=5.6\times10^{5}$ & $N=0$ \\
 & \Fri\, [SFU s] & $N=67$, Med$=4.3\times10^{6}$ & $N=18$, Med$=6.4\times10^{7}$ & $N=3$, Med$=1.7\times10^{8}$ & $N=0$ \\
\hline
$(-5.097,-4.959]$ & \Bdur\, [min] & $N=56$, Med$=12.5$ & $N=36$, Med$=21.0$ & $N=3$, Med$=32.0$ & $N=0$ \\
 & \Frp\, [SFU] & $N=56$, Med$=1.3\times10^{5}$ & $N=36$, Med$=5.5\times10^{5}$ & $N=3$, Med$=2.5\times10^{5}$ & $N=0$ \\
 & \Fri\, [SFU s] & $N=55$, Med$=1.9\times10^{7}$ & $N=36$, Med$=7.0\times10^{7}$ & $N=3$, Med$=5.6\times10^{7}$ & $N=0$ \\
\hline
$(-5.302,-4.959]$ & \Bdur\, [min] & $--$ & $--$ & $N=6$, Med$=28$ & $N=0$ \\
 & \Frp\, [SFU] & $--$ & $--$ & $N=6$, Med$=3.0\times10^{5}$ & $N=0$ \\
 & \Fri\, [SFU s] & $--$ & $--$ & $N=6$, Med$=6.6\times10^{7}$ & $N=0$ \\
\hline
${\bf (-4.959,-4.745]}$ & \Bdur\, [min] & $N=38$, Med$=14.0$ & $N=42$, Med$=20.5$ & $N=10$, Med$=44.5$ & $N=3$, Med$=58.0$ \\
 & \Frp\, [SFU] & $N=38$, Med$=1.0\times10^{5}$ & $N=42$, Med$=3.7\times10^{5}$ & $N=10$, Med$=1.3\times10^{6}$ & $N=3$, Med$=2.8\times10^{6}$ \\
 & \Fri\, [SFU s] & $N=37$, Med$=1.9\times10^{7}$ & $N=42$, Med$=7.9\times10^{7}$ & $N=10$, Med$=4.4\times10^{8}$ & $N=3$, Med$=7.4\times10^{8}$ \\
\hline
${\bf (-4.745,-4.387]}$ & \Bdur\, [min] & $N=26$, Med$=14.5$ & $N=42$, Med$=22.5$ & $N=19$, Med$=34.0$ & $N=4$, Med$=44.5$ \\
 & \Frp\, [SFU] & $N=26$, Med$=1.0\times10^{5}$ & $N=42$, Med$=1.6\times10^{5}$ & $N=19$, Med$=1.0\times10^{6}$ & $N=4$, Med$=5.8\times10^{6}$ \\
 & \Fri\, [SFU s] & $N=26$, Med$=1.6\times10^{7}$ & $N=42$, Med$=3.5\times10^{7}$ & $N=19$, Med$=1.5\times10^{8}$ & $N=4$, Med$=9.6\times10^{8}$ \\
\hline
${\bf (-4.387,-2.824]}$ & \Bdur\, [min] & $N=10$, Med$=20.0$ & $N=40$, Med$=20.5$ & $N=41$, Med$=45.0$ & $N=26$, Med$=50.5$ \\
 & \Frp\, [SFU] & $N=10$, Med$=1.0\times10^{5}$ & $N=40$, Med$=3.5\times10^{5}$ & $N=41$, Med$=2.1\times10^{6}$ & $N=26$, Med$=1.2\times10^{7}$ \\
 & \Fri\, [SFU s] & $N=9$, Med$=1.9\times10^{7}$ & $N=40$, Med$=6.6\times10^{7}$ & $N=41$, Med$=4.7\times10^{8}$ & $N=26$, Med$=2.1\times10^{9}$ \\
\enddata
\end{deluxetable}
\renewcommand{\arraystretch}{1.0}


\clearpage

\section{{ Mean coefficients for the selected probability estimation models}} \label{app:model_coefficients}
{ Table~\ref{tab:app_beta_binary} and Table~\ref{tab:app_beta_multiclass} give the mean coefficients and their uncertainty for SXR+Type-III and scale-free models. The values are estimated as described in Sect.~\ref{sec:predictive}. Note that all physical parameters are transformed to $\log_{10}$ space
before standardization, while the Boolean gate $g$ remains
untransformed. The model coefficients are transformed back to the original
logarithmic feature space before computing their mean and
standard deviation.} 
\begin{table*}[ht!]
\centering
\caption{Mean model coefficients for the selected Sun-as-a-star
SXR+Type-III and scale-free models for the confined/CME probability
problem. The reported values are $\overline{\beta}\pm\sigma_\beta$.
Here, $\beta_0$ is the model intercept, while the remaining coefficients
correspond to the respective components of the flare vector.}
\label{tab:app_beta_binary}

\setlength{\tabcolsep}{12pt}
\renewcommand{\arraystretch}{1.15}

\begin{tabular}{lll}
\hline
\hline
Model & Parameter & $\overline{\beta}\pm\sigma_\beta$ \\
\hline
SXR + Type-III
 & $\beta_0$ & $9.125\pm1.053$ \\
 & $\log F_{\rm X,p}$ & $2.591\pm0.206$ \\
 & $\log t_{\rm rise}$ & $1.471\pm0.189$ \\
 & $g$ & $-2.389\pm0.637$ \\
 & $g\log F_{\rm r,p}$ & $0.105\pm0.198$ \\
 & $g\log F_{\rm r,int}$ & $0.176\pm0.215$ \\
 & $g\log\Delta T_{\rm burst}$ & $1.918\pm0.304$ \\
\hline
Scale-free
 & $\beta_0$ & $23.225\pm1.585$ \\
 & $g$ & $-1.560\pm0.563$ \\
 & $g\log(F_{\rm r,p}/F_{\rm X,p})$ & $0.335\pm0.055$ \\
 & $g\log(\Delta T_{\rm burst}/t_{\rm rise})$ & $0.252\pm0.166$ \\
 & $\log R_{\rm X}$ & $3.141\pm0.194$ \\
\hline
\end{tabular}
\end{table*}

\begin{table*}[ht!]
\centering
\caption{Mean model coefficients for the selected Sun-as-a-star
SXR+Type-III and scale-free models for the confined/CME/SEPE probability
problem. Entries for each event class give
$\overline{\beta}\pm\sigma_\beta$; the parameter definitions are the
same as in Table~\ref{tab:app_beta_binary}.}
\label{tab:app_beta_multiclass}

\small
\setlength{\tabcolsep}{7pt}
\renewcommand{\arraystretch}{1.15}

\begin{tabular}{llccc}
\hline
\hline
Model & Parameter & Confined & CME & SEPE \\
\hline
SXR + Type-III
 & $\beta_0$
 & $-6.974\pm0.800$
 & $1.827\pm0.539$
 & $5.146\pm0.914$ \\

 & $\log F_{\rm X,p}$
 & $-2.513\pm0.166$
 & $0.031\pm0.110$
 & $2.482\pm0.195$ \\

 & $\log t_{\rm rise}$
 & $-2.098\pm0.180$
 & $-0.644\pm0.161$
 & $2.742\pm0.281$ \\

 & $g$
 & $6.053\pm0.647$
 & $4.126\pm0.586$
 & $-10.179\pm0.988$ \\

 & $g\log F_{\rm r,p}$
 & $-0.346\pm0.149$
 & $-0.292\pm0.160$
 & $0.638\pm0.095$ \\

 & $g\log F_{\rm r,int}$
 & $-0.261\pm0.160$
 & $-0.051\pm0.160$
 & $0.312\pm0.076$ \\

 & $g\log\Delta T_{\rm burst}$
 & $-2.574\pm0.286$
 & $-0.905\pm0.328$
 & $3.479\pm0.535$ \\

\hline

Scale-free
 & $\beta_0$
 & $-21.995\pm1.239$
 & $-0.337\pm0.800$
 & $22.332\pm1.368$ \\

 & $g$
 & $3.737\pm0.470$
 & $2.867\pm0.402$
 & $-6.604\pm0.650$ \\

 & $g\log(F_{\rm r,p}/F_{\rm X,p})$
 & $-0.505\pm0.047$
 & $-0.226\pm0.038$
 & $0.731\pm0.064$ \\

 & $g\log(\Delta T_{\rm burst}/t_{\rm rise})$
 & $0.497\pm0.162$
 & $0.730\pm0.156$
 & $-1.228\pm0.263$ \\

 & $\log R_{\rm X}$
 & $-3.064\pm0.153$
 & $-0.104\pm0.105$
 & $3.168\pm0.178$ \\
\hline
\end{tabular}
\end{table*}
\end{document}